\documentclass[aip,jcp,preprint,amssymb,superscriptaddress,notitlepage,showpacs]{revtex4-1}
\usepackage{graphicx}
\usepackage{amsmath}

\usepackage{color}
\usepackage{enumitem}
\usepackage{graphicx}
\usepackage[normalem]{ulem}
\usepackage{comment}

\DeclareMathOperator\erf{erf}
\DeclareMathOperator\erfc{erfc}

\newif\ifHighlitedChanges
\def\ifHighlitedChanges{\iffalse}
\ifHighlitedChanges
  
  \def\STRIKE#1{{\color{red}\sout{#1}}}
\else
  
  \def\STRIKE#1{\relax}
\fi

\begin{document}

\bibliographystyle{apsrev}
\title{Upside/Downside statistical mechanics of nonequilibrium Brownian motion. I. Distributions, moments, and correlation functions of a free particle}
\author{Galen T. Craven}  
\affiliation{Department of Chemistry, University of Pennsylvania, Philadelphia, PA  19104, USA} 
\author{Abraham Nitzan}
\affiliation{Department of Chemistry, University of Pennsylvania, Philadelphia, PA  19104, USA} 
\affiliation{School of Chemistry, Tel Aviv University, Tel Aviv 69978, Israel}

\begin{abstract}
Statistical properties of Brownian motion that arise by analyzing, separately, 
trajectories over which the system energy increases (upside) 
or decreases (downside) with respect to a threshold energy level, are derived.
This selective analysis
is applied to examine transport properties
of a nonequilibrium Brownian process that is coupled to multiple thermal sources characterized by different temperatures.
Distributions, moments, and correlation functions of a free particle that occur 
during upside and downside events
are investigated for energy activation and energy relaxation processes, and also for positive and negative energy fluctuations 
from the average energy.
The presented results are sufficiently general and can be applied 
without modification to standard Brownian motion.
This article focuses on the mathematical basis of this selective analysis.
In subsequent articles in this series we apply this general formalism to processes in which heat transfer between thermal reservoirs is mediated by activated rate processes that take place in a system bridging them.

\vspace{0.22cm}
\noindent This article may be downloaded for personal use only. 
Any other use requires prior permission of the author and AIP Publishing. 
This article appeared in \textit{J. Chem. Phys.} 148, 044101 (2018) and may be found at http://aip.scitation.org/doi/10.1063/1.5007854
\end{abstract}
 \maketitle
\section{Introduction  \label{sec:Introduction}}

The advent of nonequilibrium fluctuation theorems \cite{Jarzynski1997,Evans1993,Jarzynski1997,Kurchan1998,Crooks2000,Seifert2012} 
has driven significant advances in statistical physics,
underpinning the development of theories to describe nonequilibrium processes  
far from the linear response regime. \cite{Onsager1931,Seifert2012}
In parallel,
the derivation of macroscopic thermodynamic observables from stochastic dynamical equations \cite{Sekimoto1998,Seifert2005,Seifert2012,Van2013stochastic}  
has given credence to the role that trajectory-based approaches serve in formulations of thermodynamics. \cite{Sekimoto1998}
On a trajectory level, nonequilibrium thermal fluctuations drive
heat transfer \cite{Lebowitz1959,Lebowitz1967,Lebowitz1971,Sekimoto1998,Nitzan2003thermal,Segal2005prl,Lebowitz2008,Lebowitz2012,Sabhapandit2012,Dhar2015,Velizhanin2015,Esposito2016}
and electron-transfer-induced heat transport, \cite{craven16c,craven17a,craven17b,craven17e}
and thus gaining a fundamental understanding of the physical basis of these fluctuations is critical for the control of energy conversion in thermal transport devices.
Analyses of fluctuations are prevalent in the theoretical formulations of rate processes
in which transitions between locally stable states
are induced by thermal activation. \cite{Marcus1956,Marcus1964,Marcus1985,Marcus1993,rmp90,truh96,Komatsuzaki2001,dawn05a,Nitzan2006chemical,hern10a,Peters2015,craven15c} 
These fluctuations, and their corresponding thermal statistics, \cite{Chandler1987statmech} are typically treated as properties of the full ensemble 
by including
both positive and negative deviations from the average.
However, positive and negative deviations
can also be analyzed separately. As we will show, such selective analysis 
of a system undergoing
activation and relaxation events 
can lead to new kinetic information.

Selective statistical analysis 
is 
applied in  
economics and econometrics 
to predict the risk vs. reward of investments 
during times of increasing (upside) and decreasing (downside) value.
Analyzing upside and downside trends separately
presents new metrics and insight,
beyond what can be obtained
from analysis that takes into account the full data set. \cite{Greene2002econo,Ranganatham2006,Reilly2011,Sortino1991,Sortino1994,Keating2002universal,Ang2006downside} 
This separation procedure is performed using a selector (a measurable -- usually the value of an investment)  
and comparing how that selector compares to some threshold.
In economics the threshold could be, e.g., an opening price or the mean performance.
As we will show here, 
selective analysis of upside and downside trends can also be applied to physical phenomena in order to elucidate trends that are obscured 
through an analysis that takes into account all fluctuations.
There are many possible selectors in a physical system, 
and the chosen selector could in principle be any fluctuating observable.
However, in the stochastic picture commonly applied to condensed phase chemical dynamics, \cite{zwan01book,Nitzan2006chemical} the energy of the system is 
perhaps the most important selector due to its relation to activated events and state transitions. \cite{rmp90,dawn05a,craven15c,craven17c,craven17d}

\begin{figure}[t]
\includegraphics[width = 8.5cm,clip]{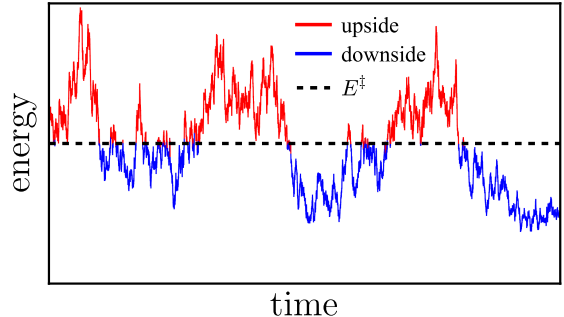}
\caption{\label{fig:upside_downside}
Energy of a representative stochastic process $x(t)$ as a function of time.
The upside segments ($E>E^\ddag$) of the trajectory 
are colored in red and the downside segments ($E<E^\ddag$)  are colored in blue.
}
\end{figure}

Heat conduction between multiple thermal sources due to vibrational interactions is a paradigmatic process in
thermal transport 
\cite{Lebowitz1959,Lebowitz1967,Lebowitz1971,Sekimoto1998,Nitzan2003thermal,Segal2005prl,Lebowitz2008,Lebowitz2012,Sabhapandit2012,Dhar2015,Velizhanin2015,Esposito2016}
and is often described using stochastic models, of which the simplest is Brownian motion. \cite{Chandrasekhar1943,OrnsteinUhlenbeck1930,zwan01book}
For a nonequilibrium Brownian stochastic process $x(t)$  that is coupled to multiple heat baths, and that is not constrained by an external potential, the energy of the system $E(t)$ depends only on a kinetic energy term that is a function
of $\dot{x}(t) = v(t)$.
Trajectories can be classified at any given time as upside or downside using the instantaneous system energy as a selector.
If the energy of the system at time $t$ is greater than a threshold energy $E^\ddag$ then the process is an upside process
at time $t$, and if the energy of the system is less than the threshold energy, the process is a downside process at time $t$. 
Obviously, the classification of a process 
as upside or downside depends on the time of analysis, and,
as shown in Fig.~\ref{fig:upside_downside}, a process can change from upside to downside and vice versa many times over the course of the trajectory.  


Here, we examine the statistical properties of a free Brownian particle that arise by
separating the full ensemble of trajectories that describe it into
upside and downside sub-ensembles with respect to various energy thresholds.
These statistical properties (moments, expectation values, correlation functions, probabilities, etc.) are classified as \textit{restricted}.
Properties that are termed \textit{unrestricted} correspond to analysis of the full ensemble.
We focus on the situation in which the full ensemble is in a steady-state,
although the developed methodology can also be applied to other cases.
The two most pertinent threshold energies are: (a) the initial energy of the process $E(0)$, which is a particular property of each individual trajectory,  and (b) the average energy $\langle E \rangle$ of the system, which is 
a statistical property of the full ensemble.
When the initial energy $E(0)$ is used, 
any calculated result is then averaged over this initial energy using the (assumed known) 
initial distribution (usually taken to be the equilibrium or a steady-state distribution). 
When $\langle E \rangle$ (a property of the initial distribution) is used, the final result is given as function of this energy. 
Using the average energy as the threshold, 
where this average corresponds to a given steady-state situation,
 has a particular meaning in statistical physics 
because of its relation to the definition of positive energy fluctuations $\delta E^+ \equiv E(t)- \langle E \rangle>0$, which are upside processes, and
negative energy fluctuations $\delta E^- \equiv E(t)- \langle E \rangle<0$, which are downside processes.
Note that $\langle E \rangle$ is a number that is the same for every trajectory while $E(0)$
is a property that depends on the initial conditions of a trajectory.

The development of an upside/downside formalism for thermalized molecular processes is principally motivated by
questions that arise with respect to what fraction of the total system energy change
and what fraction of the total heat current are contributed by each bath when a system that is driven by multiple thermal sources increases in energy and
when the system decreases in energy. Similar partitioning questions arise
with respect to positive and negative energy fluctuations from the average energy. 
However, these problems cannot be addressed using theories that treat properties of the 
full ensemble, i.e., unrestricted properties, that do not differentiate between activation and relaxation.
For example, given a system that is connected to $N$ thermal baths, each with a respective temperature $T_k$
and energy relaxation rate into the bath $\gamma_k$, applying the upside/downside formalism yields the result that
when the system energy increases or decreases by a factor $\Delta E$
the fraction of this energy that is obtained/released by each bath is
$\gamma_k T_k / \sum_k^N \gamma_k T_k$. 
While this result is intuitively plausible, its derivation is not possible using an analysis that treats the
full ensemble without separating it into upside and downside sub-ensembles.
In future articles in this series we 
apply the framework developed here to examine 
the way energy transfer between different thermal reservoirs is affected by an activated rate process in the system bridging them.
We find that in the heat transport equations that arise in  
such analysis,
dynamical transport properties appear that depend not only on the time $t$ where the upside/downside constraint is imposed, but also
on all times $t'<t$ given that that process is classified as upside or downside at $t$.
Thus, developing both one-time transport properties which depend only on time $t$ and two-time transport properties which depend on times
$t'$ and $t$ is imperative in obtaining solutions to the partitioning problems.

The remainder of this article is organized as follows:
Section~\ref{sec:Brownian} contains details of the nonequilibrium Brownian process 
that we use as a model.
In Sec.~\ref{sec:unres},
unrestricted correlation functions and moments of this process 
are derived.
The primary motivation for including the derivation of unrestricted properties is to provide a basis for counterpose with respect to the
restricted dynamical properties examined in Section~\ref{sec:res}, 
but the general integral forms of these properties can also be applied to restricted transport.
Statistical properties of restricted transport are
derived in Sec.~\ref{sec:res}.
Concluding remarks and outlook for future articles in this series are presented in Sec.~\ref{sec:conc}.

\section{Brownian Motion Driven by $N$ Thermal Sources  \label{sec:Brownian}}

The 
stochastic 
process we consider
in order to examine restricted transport and energy fluctuations 
is a free
Brownian particle that is driven by $N$ thermal sources,
where each source $k \in \left\{1,\ldots,N\right\}$ has a respective temperature $T_k$.  \cite{OrnsteinUhlenbeck1930, zwan01book} 
The Langevin equation of motion for this nonequilibrium system is 
\begin{equation}
\label{eq:EoM1}
\ddot x= - \sum_k^N\gamma_k \dot x + \sum_k^N \xi_k(t),
\end{equation}
where $\gamma_k$ is a Markovian dissipative (friction) parameter of the respective bath
and $\xi_k(t)$ is a stochastic noise term 
that obeys the correlation relations:
\begin{equation}
\begin{aligned}
 \big\langle \xi_k(t) \xi_k(t')\big\rangle &= 2 \gamma_k  k_\text{B} T_k m^{-1} \delta(t-t'), \\[0ex]
  \big\langle \xi_k(t) \xi_l(t')\big\rangle &= 0,  \\[0ex]
	\big\langle \xi_k(t) \xi_l(t)\big\rangle &= 0,  \\[0ex]
 \big\langle \xi_k(t)\big\rangle &=0, \\[0ex]
\end{aligned}
\end{equation}
for unrestricted transport,
where $\langle \ldots \rangle$ denotes an average over realizations of the noise.
The formal solution of Eq.~(\ref{eq:EoM1}) is
\begin{align}
\label{eq:xsol}
x(t) &= x(0) + \int_0^t v(s) \,ds, \\[1ex]
\label{eq:vsol}
v(t) &=  v(0) \prod^N_k e^{- \gamma_k t}  + \sum^N_l \int_0^t \Bigg(\prod^N_k e^{-\gamma_k(t-s)}\Bigg)\xi_l(s)\,ds .
\end{align}
The most common case is a process driven by two thermal sources ($N=2$), and this system has been the focus of
intensive investigation 
due to its relevance for vibrational heat conduction \cite{Lebowitz1959,Lebowitz1967,Lebowitz1971,Lebowitz2008,
Lebowitz2012,Sekimoto1998,Nitzan2003thermal,Segal2005prl,Lebowitz2012,Sabhapandit2012,Dhar2015,Velizhanin2015,Esposito2016}
and electron-transfer-induced heat transport. \cite{craven16c,matyushov16c,craven17a,craven17b,craven17e} 
In this article, 
the general derivations of the restricted properties are valid for arbitrary $N$. For simplicity, the results shown in all figures are for the $N = 2$ model.

The equation of motion (\ref{eq:EoM1}) can also be written
in a simplified form as
\begin{equation}
\label{eq:eom}
 \ddot x=
-\gamma \dot x
+ \xi(t),
\end{equation}
in terms of the
total friction and stochastic noise
\begin{equation}
\gamma = \sum_k^N \gamma_k \quad \text{and} \quad \xi(t) = \sum_k^N \xi_k(t).
\end{equation}
For convenience we will most commonly use this simplified notation.
These friction and noise terms satisfy a fluctuation-dissipation theorem
\begin{equation}
\begin{aligned}
\label{eq:noisereduced}
 \big\langle \xi(t) \xi(t')\big\rangle &= 2 \gamma k_\text{B} T m^{-1} \delta(t-t'), \\[0ex]
 \big\langle \xi(t)\big\rangle &=0.
\end{aligned}
\end{equation}
that defines an effective temperature
\begin{equation}
T = \sum_k^N \frac{\gamma_k T_k}{\gamma}.
\end{equation}
The effective inverse thermal energy is $\beta = 1 / k_\text{B} T$, where $k_\text{B}$ is the Boltzmann constant.
The general solution of Eq.~(\ref{eq:eom}) is 
\begin{align}
\label{eq:xsolsimp}
x(t) &= x(0) + \int_0^t v(s) \,ds, \\[0ex]
\label{eq:vsolsimp}
v(t) &= v(0) e^{-\gamma t}+ \int_0^t e^{-\gamma(t-s)}\xi(s)\,ds.
\end{align}
It is important to note that Eq.~(\ref{eq:eom}) is equivalent to the Langevin equation of motion for 
an equilibrium Brownian process, and therefore
all the restricted observables derived herein are also applicable to standard Brownian motion.
In future articles in the series, we apply the formalism in Eq.~(\ref{eq:EoM1}) in order to analyze specific nonequilibrium transport properties.

The Brownian process is initially characterized by a velocity distribution $\rho_0$,
and in the limit $t \to \infty$ it approaches a nonequilibrium steady state (ss) with temperature $T$.
At steady state, the velocity distribution
is a Gibbs distribution \cite{Lebowitz1959} given by
\begin{equation}
\rho^{(\text{ss})}(v) = \frac{1}{Z^{(\text{ss})}}e^{-\beta \frac{1}{2} m v^2},
\end{equation}
where $Z^{(\text{ss})}$
is the standard partition function. 
In a likely special case the initial ($t = 0$) velocity distribution of the process is the steady-state distribution:  $\rho_0 = \rho^{(\text{ss})}$.

\section{\label{sec:unres}Unrestricted Transport Properties}

In this section unrestricted statistical properties for a Brownian process that is driven by $N$ thermal sources are reviewed.
These unrestricted properties will be used in the construction of the restricted properties that are derived in later sections,
and will serve as a basis when comparing unrestricted and restricted transport. 

The transition probability for a Brownian process satisfying Eq.~(\ref{eq:eom}) is \cite{Nitzan2006chemical}
\begin{equation}
\label{eq:transprob}
\nonumber \rho\left(v\, t\, | \, v'\, t'\right) = \sqrt{\frac{1}{2 \pi \sigma_v^2(t-t')}}
\exp\left[-\Bigg(\frac{v - v' e^{-\gamma (t-t')}}{\sqrt{2 \sigma_v^2(t-t')}}\Bigg)^2\right],
\end{equation}
where
\begin{equation}
\sigma_v^2(t-t') = \frac{k_\text{B} T}{m}\Big(1-e^{-2 \gamma (t-t')}\Big),
\end{equation}
is a time-dependent variance.
Equation~(\ref{eq:transprob}) expresses the conditional probability that a process with velocity $v'$ at time $t'$  -- $\rho \left(v\, t'\, | \, v'\, t'\right) = \delta(v-v')$ -- has velocity $v$ at time $t$.
Without loss of generality the initial time can be defined as $t' = 0$, and in this case $\rho \left(v\, t\, | \, v_0\, 0\right)$
is the probability that a process with initial velocity $v(0)=v_0$ has velocity $v$ at time $t$.
The instantaneous energy of the system is $E(v) =  \tfrac{1}{2} m v^2$ and its initial value is $E(0) = \tfrac{1}{2} m v_0^2$.

The first velocity moment for a Brownian process evolving through Eq.~(\ref{eq:eom}) is
\begin{equation}
\label{eq:v}
\big\langle v(t) \big\rangle = \big\langle v(0)\big\rangle e^{-\gamma t}+ \int_0^t e^{-\gamma(t-s)}\big\langle \xi(s)\big\rangle\,ds, 
\end{equation}
and using $\big\langle \xi(s)\big\rangle =0$ from Eq.~(\ref{eq:noisereduced}), for unrestricted transport:
\begin {equation}
\big\langle v(t) \big\rangle= \big\langle v(0)\big\rangle e^{-\gamma t}.
\end{equation}
The general form of the second velocity moment is:
\begin{equation}
\begin{aligned}
\label{eq:vsqcorr}
\big\langle v^2(t)\big\rangle &= \left\langle v^2(0) \right\rangle e^{-2 \gamma t}+2   e^{-2 \gamma t} \int_0^t e^{\gamma s_1} \big\langle v(0) \xi(s_1) \big\rangle\,ds_1\\[0ex]
 & \quad +  e^{-2 \gamma t} \int_0^t \int_0^t e^{\gamma (s_1+s_2)}\big\langle\xi(s_1)\xi(s_2) \big\rangle\,ds_1\,ds_2,
\end{aligned}
\end{equation}
which can be evaluated using 
the statistical properties in
Eq.~(\ref{eq:noisereduced})
yielding
\begin{equation}
\label{eq:vsqcorr2}
\big\langle v^2(t)\big\rangle = \left\langle v^2(0) \right\rangle e^{-2 \gamma t}+ \frac{k_\text{B} T}{m} \Big(1- e^{-2 \gamma t}\Big).
\end{equation}
which is valid for $t>0$.
For a Gibbs distribution of initial velocities (or as $t \to \infty$ for any bounded initial distribution),
Eq.~(\ref{eq:vsqcorr2}) reduces to 
\begin{equation}
\label{eq:vsqunrestricted}
\big\langle v^2 \big\rangle = \frac{k_\text{B} T} {m}.
\end{equation}
In this case the system will be in a quasi-equilibrium state at the \textit{effective} temperature $T$.
The general expression for the two-time velocity correlation function is: \cite{Cohen2015review}
\begin{align}
\label{eq:vsqcorrvsst}
 \big\langle v'(t') v(t)\big\rangle &=\left\langle v^2(0) \right\rangle e^{-\gamma (t+t')}\\[0ex]
\nonumber & \quad +  e^{-\gamma (t+t')} \int_0^{t'} e^{\gamma s_1} \big\langle v(0) \xi(s_1) \big\rangle\,ds_1 \\[0ex]
\nonumber & \quad + e^{-\gamma (t+t')} \int_0^t e^{\gamma s_1} \big\langle v(0) \xi(s_1) \big\rangle\,ds_1 \\[0ex]
\nonumber & \quad + e^{-\gamma (t+t')} \int_0^t \int_0^{t'} e^{\gamma (s_1+s_2)}\big\langle\xi(s_1)\xi(s_2) \big\rangle\,ds_1\,ds_2,
\end{align}
which after applying the unrestricted noise correlations and evaluating the integrals leads to
\begin{equation}
\begin{aligned}
&\big\langle v'(t') v(t)\big\rangle =  \left\langle v^2(0) \right\rangle e^{-\gamma (t+t')} + \frac{k_\text{B} T}{m} \left(e^{-\gamma |t-t'|}- e^{-\gamma (t+t')} \right),
\end{aligned}
\end{equation}
where for notational convenience $v'\equiv v(t')$.

The average energy of a Brownian process is, 
\begin{align}
\label{eq:energy}\big\langle E(t)\big\rangle  &= \frac{1}{2} m \big\langle v^2(t)\big\rangle,
\end{align}
and for unrestricted transport under steady-state conditions,
\begin{equation}
\label{eq:avgenergy}
\langle E\rangle  = \frac{1}{2}k_\text{B} T,
\end{equation}
which illustrates that the energy of a  particle coupled to $N$ thermal reservoirs
obeys an analogous equipartition theorem with respect to the effective temperature as does a particle coupled to a single reservoir.

\section{\label{sec:res}Restricted Transport Properties}

Restricted transport properties of a free Brownian particle are derived
in this section. Details of these derivations can be found in 
the Supplementary Material.
These properties are applied in the next article in this series
to resolve questions related to energy partitioning 
in nonequilibrium processes that are driven by multiple thermal reservoirs with different temperatures, 
however, the developed formalism is sufficiently general that, apart from heat transfer
properties that are unique to the nonequilibrium situation, it can be applied
without modification to standard Brownian processes. 
As a consequence, the upside/downside mathematical framework has direct ramifications and applications 
in the theoretical formulation of 
equilibrium statistical mechanics, 
specifically analyses of fluctuations.

\subsection{\label{sec:resttrans}Restricted Transition Probabilities and Distributions}


The general forms for the conditional probability that 
the energy of a process is either above (upside process  $\uparrow$) or below (downside process $\downarrow$) the threshold energy $E^\ddag$ at time $t$ given that the system was characterized by the initial distribution $\rho_0$ at $t = 0$ are:
\begin{align}
\label{eq:probupGibbsthres}
p_\uparrow \Big(t\,\big|\,E(t)>E^\ddag, \rho_0 \,0 \Big) &= \int_{\mathbb{R}^2} \rho_0(\bar{v}) \rho(v\, t\, | \, \bar{v}\, 0) \Theta\big(E(v)-E^\ddag\big) dv \,d\bar{v},\\[1ex]
\label{eq:probdownGibbsthres}
p_\downarrow \Big(t\,\big|\,E(t)<E^\ddag, \rho_0 \,0 \Big) &=  \int_{\mathbb{R}^2} \rho_0(\bar{v}) \rho(v\, t\, | \, \bar{v}\, 0) \Theta\big(E^\ddag-E(v)\big) dv \,d\bar{v}.
\end{align}
where $\Theta$ is the Heaviside function and $\mathbb{R}^n$ denotes integration over $n$-dimensional real space.
These restricted probabilities can be applied to construct the 
restricted probability densities.
The general forms for the restricted densities 
of $v$ at time $t$ given that the system was initially characterized by velocity distribution $\rho_0$ are
\begin{align}
\label{eq:transprobnegupdistthres}
\tilde{\rho}_\uparrow \big(v\, t\, | \, \rho_0 \, 0\big) &= \frac{ \displaystyle \int_{\mathbb{R}}  
\rho_0(\bar{v}) \rho(v\, t\, | \, \bar{v}\, 0)\Theta\big(E(v)-E^\ddag\big) d\bar{v}}
{\displaystyle p_\uparrow \Big(t\,\big|\,E(t)>E^\ddag, \rho_0 \,0 \Big)}, \\[1ex]
\label{eq:transprobnegdowndistthres}
\tilde{\rho}_\downarrow\big(v\, t\, | \, \rho_0 \, 0\big) &=  \frac{ \displaystyle \int_{\mathbb{R}}   
\rho_0(\bar{v}) \rho(v\, t\, | \, \bar{v}\, 0)\Theta\big(E^\ddag-E(v)\big)  d\bar{v}}
{\displaystyle p_\downarrow \Big(t\,\big|\,E(t)<E^\ddag, \rho_0 \,0 \Big)}.
\end{align}
where $\tilde{\rho}$ represents a restricted density.
From these general expressions,  the restricted transition probability densities for different thresholds can be derived.

\begin{figure*}[t]
\includegraphics[width = 17.0cm,clip]{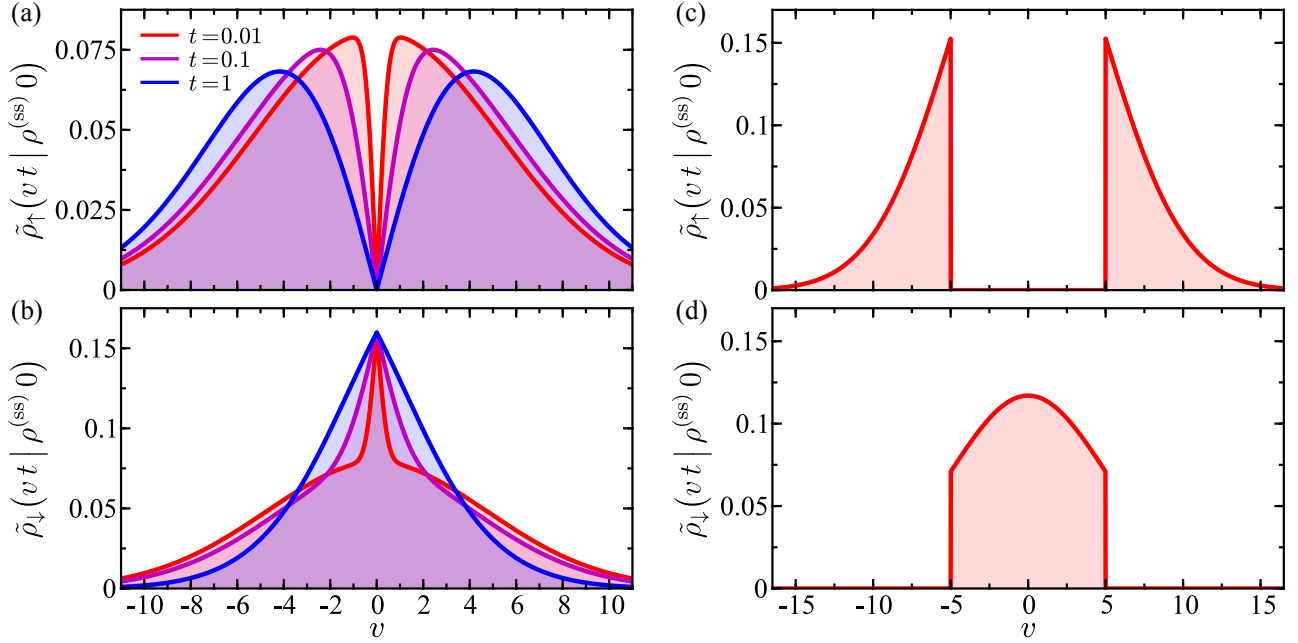}
\caption{\label{fig:probdens}
Restricted upside and downside probability density for $\rho_0 = \rho^{(\text{ss})}$ with (a)-(b) $E^\ddag = E(0)$ 
and (c)-(d) $E^\ddag = \left\langle E\right\rangle$.
The time-dependent densities in (a)-(b) are evaluated at different times marked in the legend of (a). 
Parameters in this and all other figures are $\gamma = 1$ ($\gamma_1 = 1/4$, $\gamma_2 = 3/4$), $m = 1$, and $T = 1$ ($T_1 = 4/5$, $T_2 = 16/15$)
which are given in reduced units with 
characteristic dimensions: length $\widetilde{\sigma} = 1\,\text{\AA}$,  time $\widetilde{\tau} = 1\,\text{ps}$,
mass $\widetilde{m} = 10\,m_u$,
and temperature $\widetilde{T} = 300\,\text{K}$.
}
\end{figure*}

\textit{$E(t)$ compared to $E(0)$}---
For $E^\ddag = E(0)$ and $\rho_0 = \rho^{(\text{ss})}$ 
(where $E(0)$ is a function of $\bar{v}$ which is averaged over the distribution $\rho_0(\bar{v})$ to obtain the expectation values), \cite{note1,note2}
evaluating Eqs.~(\ref{eq:probupGibbsthres}) and (\ref{eq:probdownGibbsthres}) gives the expected result: 
\begin{equation}
\label{eq:ssprob}
p_\uparrow \Big(t\,\big|\,E(t)>E(0), \rho^{(\text{ss})} \,0\Big) =  p_\downarrow \Big(t\,\big|\,E(t)<E(0), \rho^{(\text{ss})} \,0 \Big) = \frac{1}{2},
\end{equation}
which shows that a trajectory sampled from the steady-state distribution 
has equal probability to increase or decrease in energy over time interval $[0,t]$, and that these probabilities are time-independent
and temperature-independent.   

The restricted probability densities in $v$ are:
\begin{align}
\label{eq:transprobnegupdistss}
\tilde{\rho}_\uparrow \big(v\, t\, | \, \rho^{(\text{ss})} \, 0\big) &=  \rho^{(\text{ss})}(v) \left[\frac{ \displaystyle \Phi\big(\lvert v \rvert,v; t\big)-\Phi\big(-\lvert v \rvert,v; t\big)}{p_\uparrow \Big(t\,\big|\,E(t)>E(0), \rho^{(\text{ss})} \,0\Big) }\right],\\[1ex]
\label{eq:transprobnegdowndistss}
\tilde{\rho}_\downarrow \big(v\, t\, | \, \rho^{(\text{ss})} \, 0\big) &= \rho^{(\text{ss})}(v)\left[ \frac{ \displaystyle 1- \Phi\big(\lvert v \rvert,v; t\big)+\Phi\big(-\lvert v \rvert,v; t\big)}{p_\downarrow \Big(t\,\big|\,E(t)>E(0), \rho^{(\text{ss})} \,0\Big) }\right], 
\end{align}
where
\begin{align}
\label{eq:cdf}
 \Phi(\alpha,v_0; t) &= \int_{-\infty}^{\alpha}  \rho(v\, t\, | \, v_0\, 0)\,dv \\
	&= \frac{1}{2}\erfc\left[\frac{ v_0 e^{-\gamma t} - \alpha}{\sqrt{2 \sigma_v^2(t)}}\right],
\end{align}
is the time-dependent cumulative distribution function of the 
probability density $\rho(v\, t\, | \, v_0\, 0)$.
The upside and downside densities are shown in Figs.~\ref{fig:probdens}(a)-(b)
as functions of $v$ for various values of $t$.
In this case, the restricted densities are symmetric (even) functions in $v$ for all $t$. 
Also observe that both the upside and downside restricted probability densities 
are not Gaussian and have a singularity at $v = 0$.


\textit{$E(t)$ relative to  $\left\langle E\right\rangle$}---
A process can also be classified as upside or downside with respect to the average
energy $\langle E \rangle$ (an ensemble property) instead of the initial energy $E(0)$ (a property of each trajectory individually).
We denote a positive energy fluctuation from the average energy (upside)  as
\begin{equation}
\label{eq:flucup}
\delta E^+ \equiv E(t)- \langle E \rangle>0,
\end{equation}
and a negative energy fluctuation from the average (downside) as
\begin{equation}
\label{eq:flucdown}
\delta E^- \equiv E(t)- \langle E \rangle<0.
\end{equation}
When using this threshold the ``upside'' and ``downside'' terms simply imply that the system energy is above or below the
average value, respectively.
A likely special case is a system with energy threshold $E^\ddag = \langle E\rangle$ and initial distribution $\rho_0 = \rho^{(\text{ss})}$.
Using Eqs.~(\ref{eq:probupGibbsthres}) and (\ref{eq:probdownGibbsthres}) then leads to:
\begin{align}
\label{eq:probupGibbsthresavg}
p_\uparrow &\Big(t\,\big|\,\delta E^+, \rho^{(\text{ss})}  \,0 \Big) = \erfc{\left(\sqrt{1/2}\right)}, \\[1ex] 
\label{eq:probdownGibbsthresavg}
p_\downarrow &\Big(t\,\big|\,\delta E^-, \rho^{(\text{ss})}  \,0 \Big) = \erf{\left(\sqrt{1/2}\right)}, 
\end{align}
after evaluating Eqs.~(\ref{eq:probupGibbsthres}) and (\ref{eq:probdownGibbsthres}),
which are temperature-independent.

For positive and negative energy fluctuations from the average 
which are categorized through application of $E^\ddag =  \left\langle E \right\rangle$ as the energy threshold
the restricted densities in $v$ are
\begin{align}
\label{eq:transprobnegupdistthresfluc}
\tilde{\rho}_\uparrow \big(v\, t\, | \, \rho^{(\text{ss})} \, 0\big) &= \frac{\displaystyle \rho^{(\text{ss})}(v) \, \Theta\big(E(v)-\left\langle E \right\rangle\!\big)}{\displaystyle p_\uparrow \Big(t\,\big|\,\delta E^+, \rho^{(\text{ss})}  \,0 \Big)} , \\[1ex]
\label{eq:transprobnegdowndistthresfluc}
\tilde{\rho}_\downarrow\big(v\, t\,  | \, \rho^{(\text{ss})} \, 0\big) &= \frac{\displaystyle \rho^{(\text{ss})}(v) \, \Theta\big(\!\left\langle E \right\rangle-E(v)\big)}{\displaystyle p_\downarrow \Big(t\,\big|\,\delta E^-, \rho^{(\text{ss})}  \,0 \Big)} ,
\end{align}
which are time-independent because $p_\uparrow$ and $p_\downarrow$ are time-independent. 
In this case, the restricted densities are truncated Gaussian distributions that are normalized over the respective upside or downside region.
These distributions are shown in Figs.~\ref{fig:probdens}(c)-(d) where it can be observed that
each distribution has singularities at $v = \pm \sqrt{\langle v^2\rangle }$.
Note that the application of an ensemble-based constant energy threshold, specifically $\left\langle E \right\rangle$, results in restricted distributions 
which have differing geometrical properties than those obtained using a 
trajectory-dependent threshold $E^\ddag = E_(0)$ which are shown in Figs.~\ref{fig:probdens}(a)-(b).
Namely, when using the average system energy as a threshold the upside probability density $\tilde{\rho}_\uparrow$ is nonzero on the discontinuous interval $(-\infty,-\sqrt{\langle v^2\rangle }) \cup (\sqrt{\langle v^2\rangle },\infty)$ and the downside density
$\tilde{\rho}_\downarrow$ is nonzero on the continuous interval $(-\sqrt{\langle v^2\rangle }, \sqrt{\langle v^2\rangle })$.
Also observe that the restricted densities obtained using a constant energy threshold are symmetric.

\subsection{\label{sec:restvel}Restricted Moments: Velocity}

The restricted probabilities and probability densities derived in Sec.~\ref{sec:resttrans} can be used to
construct the restricted velocity moments, which, for Brownian motion, are proportional to the restricted energy moments.
These moments appear in the expressions for the restricted heat currents that are investigated in the next article in this series,
and thus have direct implications for upside/downside thermal transport and energy partitioning.

In explicit integral form, the $k$th restricted raw moments of $v$ for threshold $E^\ddag$ given that the system was initially characterized 
by distribution $\rho_0$ are
\begin{align}
\nonumber \Big\langle  v^k\big(t\,\big|\,E(t)>E^\ddag, \rho_0 \,0 \big)\Big\rangle_\uparrow &\equiv \left\langle  v^k(t)\right\rangle_\uparrow   \\ 
&= 
\frac{ \displaystyle \int_{\mathbb{R}^2}  v^k  
\rho_0(\bar{v}) \rho(v\, t\, | \, \bar{v}\, 0) \Theta\big(E(v)-E^\ddag\big) dv \,d\bar{v}}
{\displaystyle p_\uparrow \Big(t\,\big|\,E(t)>E^\ddag, \rho_0 \,0 \Big)},\\[1ex]
\nonumber \Big\langle v^k\big(t\,\big|\,E(t)<E^\ddag,  \rho_0 \,0 \big)\Big\rangle_\downarrow &\equiv \left\langle  v^k(t)\right\rangle_\downarrow \\
&=   
\frac{ \displaystyle \int_{\mathbb{R}^2} v^k  
\rho_0(\bar{v}) \rho(v\, t\, | \, \bar{v}\, 0) \Theta\big(E^\ddag - E(v)\big) dv \,d\bar{v}}
{\displaystyle p_\downarrow \Big(t\,\big|\,E(t)<E^\ddag, \rho_0 \,0 \Big)}.
\end{align}
where the numerator in each expression is a normalization factor.

\begin{figure}[t]
\includegraphics[width = 8.5cm,clip]{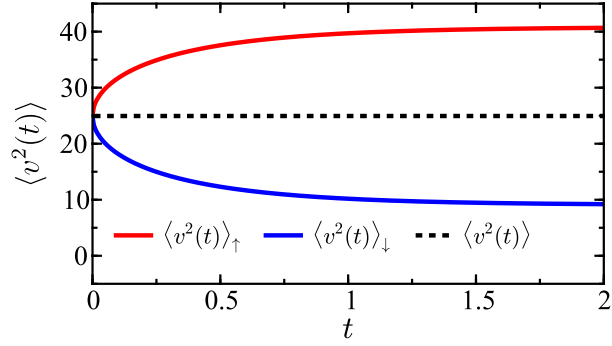}
\caption{\label{fig:velsq_ss}
Velocity moment $\langle v^2(t) \rangle$ as a function of $t$
for restricted (solid) and unrestricted (dashed) transport 
and a steady-state initial distribution $\rho^{(\text{ss})}$.
The threshold energy is $E^\ddag = E(0)$.
}
\end{figure}

\textit{$E(t)$ compared to $E(0)$}---
Under steady-state conditions,
\begin{align}
 \Big\langle  v\big(t\,\big|\,E(t)>E(0),  \rho^{(\text{ss})} \,0 \big)\Big\rangle_\uparrow    &= 0, \\[1ex]
\Big\langle v\big(t\,\big|\,E(t)<E(0), \rho^{(\text{ss})} \,0 \big)\Big\rangle_\downarrow   &=  0. 
\end{align}
as expected from the corresponding symmetrical restricted probability densities (see  Figs.~\ref{fig:probdens}(c)-(d)).
The second restricted moments are,
\begin{align}
\label{eq:vsqssup}
&\Big\langle  v^2\big(t\,\big|\,E(t)>E(0), \rho^{(\text{ss})} \,0 \big)\Big\rangle_\uparrow  
=  \frac{k_\text{B} T} { m} \bigg[1+\frac{2}{\pi} G(t) \bigg],\\[1ex]
\label{eq:vsqssdown}
&\Big\langle v^2\big(t\,\big|\,E(t)<E(0), \rho^{(\text{ss})}\,0 \big)\Big\rangle_\downarrow 
= \frac{k_\text{B} T} { m} \bigg[1-\frac{2}{\pi} G(t) \bigg],
\end{align}
with
\begin{equation}
\label{eq:G}
G(t) = \sqrt{1-e^{-2 \gamma t}}.
\end{equation}
The unrestricted and restricted second moments are shown in Fig.~\ref{fig:velsq_ss} as functions of time where it can be observed
that after an initial transient decay the restricted moments approach respective asymptotic values
corresponding to $G(t) \to 1$ in Eqs.~(\ref{eq:vsqssup}) and (\ref{eq:vsqssdown}).
Figure~\ref{fig:velsq_ss} also illustrates that for threshold $E(0)$, when the full ensemble is in a steady-state, the equality
\begin{equation}
\Big| \Big\langle  v^2\big(t\,\big|\,E(t)>E(0), \rho^{(\text{ss})} \,0 \big)\Big\rangle_\uparrow  - \langle v^2(t)\rangle \Big| = \Big| \Big\langle v^2\big(t\,\big|\,E(t)<E(0), \rho^{(\text{ss})}\,0 \big)\Big\rangle_\downarrow  - \langle v^2(t)\rangle \Big|,
\end{equation}
holds which states that upside and downside second moments are split symmetrically about the unrestricted moment.

\textit{$E(t)$ relative to  $\left\langle E\right\rangle$}---
Using energy threshold $E^\ddag  = \left\langle E\right\rangle$,
the first restricted moments of the velocity for a system initially characterized by the distribution $\rho_0$ are,
\begin{align}
\nonumber \Big\langle  v\big(t\,\big|\,\delta E^+,\rho_0 \,0 \big)\Big\rangle_\uparrow &\equiv \big\langle  v(t)\big\rangle_\uparrow    \\ 
& = \frac{ \displaystyle \int_{\mathbb{R}^2}  v  
\rho_0(\bar{v}) \rho(v\, t\, \big| \, \bar{v}\, 0) \Theta\big(E(v)-\left\langle E\right\rangle\!\big)\,dv \,d\bar{v}}
{\displaystyle p_\uparrow \Big(t\,\big|\,\delta E^+, \rho_0 \,0 \Big)},\\[1ex]
\nonumber  \Big\langle v\big(t\,\big|\,\delta E^-,  \rho_0 \,0 \big)\Big\rangle_\downarrow &\equiv \big\langle  v(t)\big\rangle_\downarrow  \\ 
& = \frac{ \displaystyle \int_{\mathbb{R}^2}  v  
\rho_0(\bar{v}) \rho(v\, t\, \big| \, \bar{v}\, 0) \Theta\big(\!\left\langle E\right\rangle-E(v)\big) \,dv \,d\bar{v}}
{\displaystyle p_\downarrow \Big(t\,\big|\,\delta E^-, \rho_0 \,0 \Big)},
\end{align}
and for the special case with $\rho_0 = \rho^{(\text{ss})}$ the expected results are recovered:
\begin{align}
\Big\langle  v\big(t\,\big|\,\delta E^+,  \rho^{(\text{ss})} \,0 \big)\Big\rangle_\uparrow    &= 0,\\[1ex]
 \Big\langle v\big(t\,\big|\,\delta E^-, \rho^{(\text{ss})} \,0 \big)\Big\rangle_\downarrow  &= 0.
\end{align}
The second restricted velocity moments are, correspondingly,
\begin{align}
\nonumber \Big\langle  v^2\big(t\,\big|\,\delta E^+, \rho_0 \,0 \big)\Big\rangle_\uparrow  &\equiv  \big\langle  v^2(t)\big\rangle_\uparrow  \\
  &= 
\frac{ \displaystyle \int_{\mathbb{R}^2}  v^2  
\rho_0(\bar{v}) \rho(v\, t\, | \, \bar{v}\, 0) \Theta\big(E(v)-\left\langle E\right\rangle\!\big) \,dv \,d\bar{v}}
{\displaystyle p_\uparrow \Big(t\,\big|\,\delta E^+, \rho_0 \,0 \Big)},\\[1ex]
\nonumber \Big\langle v^2\big(t\,\big|\,\delta E^-, \rho_0 \,0 \big)\Big\rangle_\downarrow &\equiv  \big\langle  v^2(t)\big\rangle_\downarrow \\
   &= 
\frac{ \displaystyle \int_{\mathbb{R}^2}  v^2  
\rho_0(\bar{v}) \rho(v\, t\, | \, \bar{v}\, 0) \Theta\big(\!\left\langle E\right\rangle-E(v)\big) \,dv \,d\bar{v}}
{\displaystyle p_\downarrow \Big(t\,\big|\,\delta E^-, \rho_0 \,0 \Big)},
\end{align}
and 
\begin{align}
\label{eq:vsqssupfluc}
\Big\langle  v^2\big(t\,\big|\,\delta E^+, \rho^{(\text{ss})} \,0 \big)\Big\rangle_\uparrow   & = \frac{k_\text{B} T} {m}\left[1+\sqrt{\frac{2}{\pi e}} \left(\frac{ 1}{\erfc{(\sqrt{1/2})}}\right)\right]\approx 2.53 \times \frac{k_\text{B} T} {m},\\[1ex]
\label{eq:vsqssdownfluc}
\Big\langle v^2\big(t\,\big|\,\delta E^-, \rho^{(\text{ss})} \,0 \big)\Big\rangle_\downarrow  & = \frac{k_\text{B} T} {m}\left[1-\sqrt{\frac{2}{\pi e}} \left(\frac{ 1}{\erf{(\sqrt{1/2})}}\right)\right]\approx 0.291 \times  \frac{k_\text{B} T} {m}.
\end{align}
The time-independence of the 
restricted moments for this energy threshold is a direct consequence of stationarity in the corresponding probability densities (see Fig.~\ref{fig:probdens}(c)-(d)).

\subsection{\label{sec:restenergy}Restricted Moments: Energy}

The expectation value of the system energy $E$ at time $t$ in the restricted spaces
corresponding to upside and downside processes can be calculated directly using the respective restricted second velocity moment derived in Sec.~\ref{sec:restvel}. 
The general expressions for the restricted energy expectation values at time $t$
given that the system was initially characterized by the distribution $\rho_0$ are
\begin{align}
\label{eq:Eupdist}
 \nonumber \Big\langle  E\big(t\,\big|\,E(t)>E^\ddag, \rho_0 \,0\big)\Big\rangle_\uparrow    &=  \frac{1}{2}m\Big\langle v^2\big(t\,\big|\,E(t)>E^\ddag \rho_0 \,0 \big)\Big\rangle_\uparrow,  \\[1ex]
\nonumber \Big\langle E\big(t\,\big|\,E(t)<E^\ddag,  \rho_0 \,0\big)\Big\rangle_\downarrow   &=  \frac{1}{2}m \Big\langle v^2\big(t\,\big|\,E(t)<E^\ddag,  \rho_0 \,0 \big)\Big\rangle_\downarrow, 
\end{align}
where the upside and downside processes are separated using energy threshold $E^\ddag$.

\subsection{Two-time Restricted Transition Probabilities and Distributions}

Up to this point we have only considered the statistical properties of a process at time $t$ given that the process is upside/downside at time $t$. 
A more general analysis that includes future- or history-dependence 
can be performed by observing fluctuations at time $t'>t$ or $t'<t$ given that the process is upside/downside at time $t$. In other words,
a process can be selected as upside or downside at time $t$ and then the statistical properties along that process 
at some time $t'>t$ or $t'<t$ can be constructed.
This two-time analysis will be used to evaluate two-time correlation functions and transport properties,
and to resolve the question of what fraction of the total energy change and total heat current is contributed by each bath
during upside and downside events. 
In what follows we limit ourselves to the case of $t'<t$.
The reason is that in future articles in this series,
integrals appear in the heat transport equations which contain energy fluxes from each bath as integrands that must be calculated at time $t'$ from the group of trajectories that are upside/downside at future time $t>t'$. 
Applying this two-time selective analysis in the limiting case of $t' = t$ will recover the properties derived in Sections~\ref{sec:resttrans} and \ref{sec:restvel}.

\begin{figure}[t]
\includegraphics[width = 8.5cm,clip]{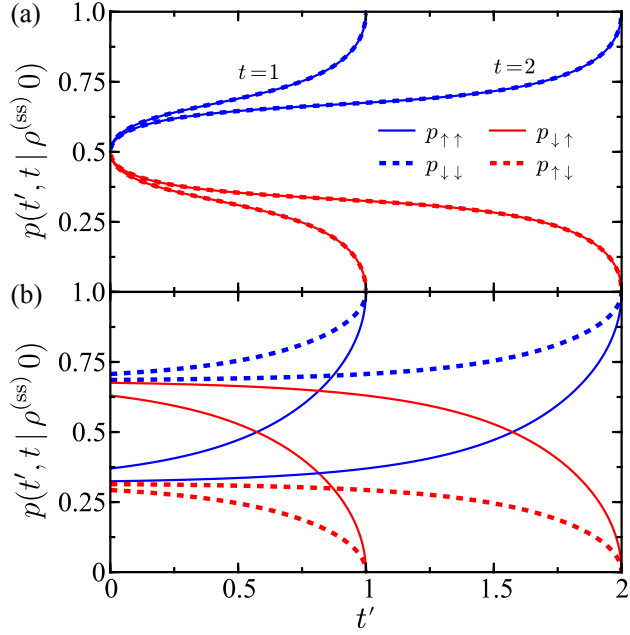}
\caption{\label{fig:transprobtimetwotime}
Two-time twice-restricted probabilities
$p_{\uparrow\uparrow}$ (blue; solid), 
$p_{\downarrow\downarrow}$ (blue; dashed),
$p_{\downarrow\uparrow}$ (red; solid), and 
$p_{\uparrow\downarrow}$ (red; dashed)
as a function of $t'$ for $t = 1$ and $t = 2$
and energy thresholds (a) $E(0)$
and (b) $\left\langle E\right\rangle$.
In both panels $\rho_0 = \rho^{(\text{ss})}$.
}
\end{figure}

The first two-time observable we examine is the conditional probability that a process that is upside/downside at time $t$ is also upside/downside at time $t'<t$.
There are four possible combinations of upside and downside events:
upside at both $t'$ and $t$ ($\uparrow\uparrow$), 
downside at both $t'$ and $t$ ($\downarrow\downarrow$),
downside at $t'$ and upside at $t$ ($\downarrow\uparrow$),
upside at $t'$ and downside at $t$ ($\uparrow\downarrow$).
Calculation of these probabilities involves evaluation of transition probabilities with constraints at both time $t'$ and time $t$.
The general forms for these these two-time twice-restricted upside/downside-upside/downside probabilities given that the system is initial characterized by distribution $\rho_0$ are:
\begin{align}
 \nonumber p_{\uparrow\uparrow} \Big(t', t\,\big| \,E(t')>E^\ddag, E(t)>E^\ddag, \rho_0 \,0 \Big) \!\! &\equiv p_{\uparrow\uparrow} (t',t \,|  \, \rho_0 \,0 )  \\
\nonumber &= \int_{\mathbb{R}^3} \rho_0(\bar{v}) \rho\big(v\, t | \, v'\, t'\big) \rho\big(v'\, t'\, | \, \bar{v}\, 0\big)\\
& \nonumber \qquad \times \Theta\big(E(v')-E^\ddag\big)\Theta\big(E(v)-E^\ddag\big)\,d\bar{v}\,dv'dv \\
& \nonumber \quad  \bigg/ \!\! \int_{\mathbb{R}^3}   \rho_0(\bar{v}) \rho\big(v\, t | \, v'\, t'\big) \rho\big(v'\, t'\, | \, \bar{v}\, 0\big) \\
& \qquad \times \Theta\big(E(v) -E^\ddag \big)\,d\bar{v}\,dv'dv,
\end{align} 
\begin{align}
 \nonumber p_{\downarrow\downarrow} \Big(t', t\,\big| \,E(t')<E^\ddag, E(t)<E^\ddag, \rho_0 \,0 \Big) \!\! &\equiv p_{\uparrow\uparrow} (t',t \,|  \, \rho_0 \,0 ) \\
\nonumber &= \int_{\mathbb{R}^3} \rho_0(\bar{v}) \rho\big(v\, t | \, v'\, t'\big) \rho\big(v'\, t'\, | \, \bar{v}\, 0\big)\\
& \nonumber \qquad \times \Theta\big(E^\ddag-E(v')\big)\Theta\big(E^\ddag-E(v)\big)\,d\bar{v}\,dv'dv \\
& \nonumber \quad  \bigg/ \!\! \int_{\mathbb{R}^3}   \rho_0(\bar{v}) \rho\big(v\, t | \, v'\, t'\big) \rho\big(v'\, t'\, | \, \bar{v}\, 0\big) \\
& \qquad \times \Theta\big(E^\ddag-E(v)\big)\,d\bar{v}\,dv'dv,
\end{align} 
\begin{align}
\nonumber p_{\downarrow\uparrow} \Big(t', t\,\big| \,E(t')<E^\ddag, E(t)>E^\ddag, \rho_0 \,0 \Big) \!\!&\equiv p_{\downarrow\uparrow} (t',t \,|  \, \rho_0 \,0 ) \\
\nonumber &= \int_{\mathbb{R}^3} \rho_0(\bar{v}) \rho\big(v\, t | \, v'\, t'\big) \rho\big(v'\, t'\, | \, \bar{v}\, 0\big)\\
& \nonumber \qquad \times \Theta\big(E^\ddag-E(v')\big)\Theta\big(E(v)-E^\ddag\big)\,d\bar{v}\,dv'dv \\
&  \nonumber \quad  \bigg/ \!\! \int_{\mathbb{R}^3} \rho_0(\bar{v})  \rho\big(v\, t | \, v'\, t'\big) \rho\big(v'\, t'\, | \, \bar{v}\, 0\big) \\
& \qquad \times  \Theta\big(E(v)-E^\ddag \big)\,d\bar{v}\,dv'dv, 
\end{align} 
\begin{align}
\nonumber p_{\uparrow\downarrow} \Big(t', t\,\big| \,E(t')>E^\ddag, E(t)<E^\ddag, \rho_0 \,0 \Big) \!\!&\equiv p_{\uparrow\downarrow} (t',t \,|  \, \rho_0 \,0 )\\
\nonumber &=\int_{\mathbb{R}^3} \rho_0(\bar{v}) \rho\big(v\, t | \, v'\, t'\big) \rho\big(v'\, t'\, | \, \bar{v}\, 0\big)\\
& \nonumber \quad \times \Theta\big(E(v')-E^\ddag\big)\Theta\big(E^\ddag - E(v) \big)\,d\bar{v}\,dv'dv \\
&  \nonumber\quad \bigg/ \!\!\int_{\mathbb{R}^3} \rho_0(\bar{v})  \rho\big(v\, t | \, v'\, t'\big) \rho\big(v'\, t'\, | \, \bar{v}\, 0\big) \\
& \qquad \times \Theta\big(E^\ddag - E(v) \big)\,d\bar{v}\,dv'dv, 
\end{align}
where 
we have applied a constrained forward Chapman-Kolmogorov evolution \cite{Nitzan2006chemical} to construct the integrands.
These integrals with multiple constraints are, in general, either algebraically cumbersome or not analytically tractable 
and therefore are most easily evaluated using
numerical procedures.

The two-time twice-restricted upside/downside-upside/downside conditional probabilities for energy thresholds $E(0)$ and $\left\langle E\right\rangle$
are shown in Figs.~\ref{fig:transprobtimetwotime}(a) and (b), respectively. 
For $E^\ddag = E(0)$ (a trajectory-dependent threshold)
the probabilities are related by the equalities $p_{\uparrow\uparrow} = p_{\downarrow\downarrow}$
and $p_{\downarrow\uparrow} = p_{\uparrow\downarrow}$ for all $t'$.
Distinct trends are observed at the limits of the $[0,t]$ time interval, 
specifically, as $t'\to 0$:
\begin{equation}
 \lim_{t'\to 0}p_{\uparrow\uparrow}=\lim_{t'\to 0}p_{\downarrow\downarrow} = \lim_{t'\to 0}p_{\downarrow\uparrow} = \lim_{t'\to 0}p_{\uparrow\downarrow} = 1/2,
\end{equation}
(consistent with Eq.~(\ref{eq:ssprob})) and at the opposite limit $t' \to t$ of the interval:
\begin{equation}
\label{eq:limitstwotimeprob}
 \lim_{t'\to t}p_{\uparrow\uparrow}= \lim_{t'\to t}p_{\downarrow\downarrow} = 1 \quad \text{and} \quad  \lim_{t'\to t}p_{\downarrow\uparrow} =  \lim_{t'\to t}p_{\uparrow\downarrow} = 0.
\end{equation}
As $t'$ approaches $t$ the probabilities increase rapidly as quantified by $(\partial p/ \partial t')_t$,
which implies that if a process is upside or downside at $t$, then 
a short time before at $t' = t - \Delta t$ the process was likely to be in the same state.
Moreover, for this specific energy threshold, the most likely outcome is that
a process that is upside or downside at $t$ was in the same
restricted state at any other $t'>0$.

When the average system energy $\left\langle E\right\rangle$ (an ensemble property that is the same for each trajectory) 
is used as the threshold, different relations between the two-time probabilities are observed
as shown in Fig.~\ref{fig:transprobtimetwotime}(b).
In this case, $p_{\uparrow\uparrow} \neq p_{\downarrow\downarrow}$ and $p_{\downarrow\uparrow}  \neq p_{\uparrow\downarrow}$,
except at the $t'= t$ point where the equalities given in
Eq.~(\ref{eq:limitstwotimeprob}) are recovered.
For this threshold, the restricted probabilities also show different temporal behavior
than for the case with threshold $E(0)$.
Namely, processes that are upside at $t$ are more likely to be in the opposite state at $t' = 0$, i.e., if 
a process is upside at $t' = t$ it was most likely downside at $t' = 0$,
and processes that are downside at $t$ are more likely to be in the same state at $t' = 0$, i.e., if 
a process is downside at $t' = t$ it was most likely downside at $t' = 0$.
This implies that the probabilities for positive and negative energy fluctuations from the average energy are dynamically heterogeneous and not symmetric
with respect to the unrestricted properties, which
is different from the case of positive and negative energy changes which are split symmetrically
about the unrestricted probability.

Next consider the restricted probability densities 
at time $t'<t$.
For a system with energy threshold $E^\ddag$
and initial distribution $\rho_0$, 
the general expression for the two-time once-restricted probability density of $v'$ 
at $t'<t$ given that the process is upside at $t$ is
\begin{align}
\label{eq:densityupbackwardsgen}
\nonumber \tilde{\rho}_\uparrow \Big(v'\,t'<t\,\big|\,E(t)>E^\ddag, \rho_0 \,0 \Big)^< &\equiv \tilde{\rho}_\uparrow\left(v'\,t'\,|\,\rho_0 \,0 \right)^< \\
&=\frac{ \displaystyle \int_{\mathbb{R}^2} \rho_0(\bar{v}) \rho\big(v\, t | \, v'\, t'\big) \rho\big(v'\, t'\, | \, \bar{v}\, 0\big)\Theta\big(E(v) - E^\ddag \big)\,d\bar{v}\,dv}{ \displaystyle \int_{\mathbb{R}^3} \rho_0(\bar{v})  \rho\big(v\, t | \, v'\, t'\big) \rho\big(v'\, t'\, | \, \bar{v}\, 0\big)\Theta\big(E(v) - E^\ddag \big)\,d\bar{v}\,dv'dv},
\end{align}
and the corresponding density at $t'$ given that the process is downside at $t$ is
\begin{align}
\label{eq:densitydownbackwardsgen}
\nonumber \tilde{\rho}_\downarrow \Big(v'\,t'<t\,\big|\,E(t)<E^\ddag, \rho_0 \,0 \Big)^<&\equiv \tilde{\rho}_\downarrow\left(v'\,t'\,|\,\rho_0 \,0 \right)^< \\
&=\frac{ \displaystyle \int_{\mathbb{R}^2}\rho_0(\bar{v}) \rho\big(v\, t | \, v'\, t'\big) \rho\big(v'\, t'\, | \, \bar{v}\, 0\big)\Theta\big(E^\ddag - E(v)\big)\,d\bar{v}\,dv}{ \displaystyle \int_{\mathbb{R}^3} \rho_0(\bar{v})  \rho\big(v\, t | \, v'\, t'\big) \rho\big(v'\, t'\, | \, \bar{v}\, 0\big)\Theta\big(E^\ddag - E(v)\big)\,d\bar{v}\,dv'dv}.
\end{align}
The superscript ``$<$'' indicates that the density is calculated at $t'<t$ while the upside/downside constraint is imposed only at future time $t$.

\begin{figure*}[t]
\includegraphics[width = 17.0cm,clip]{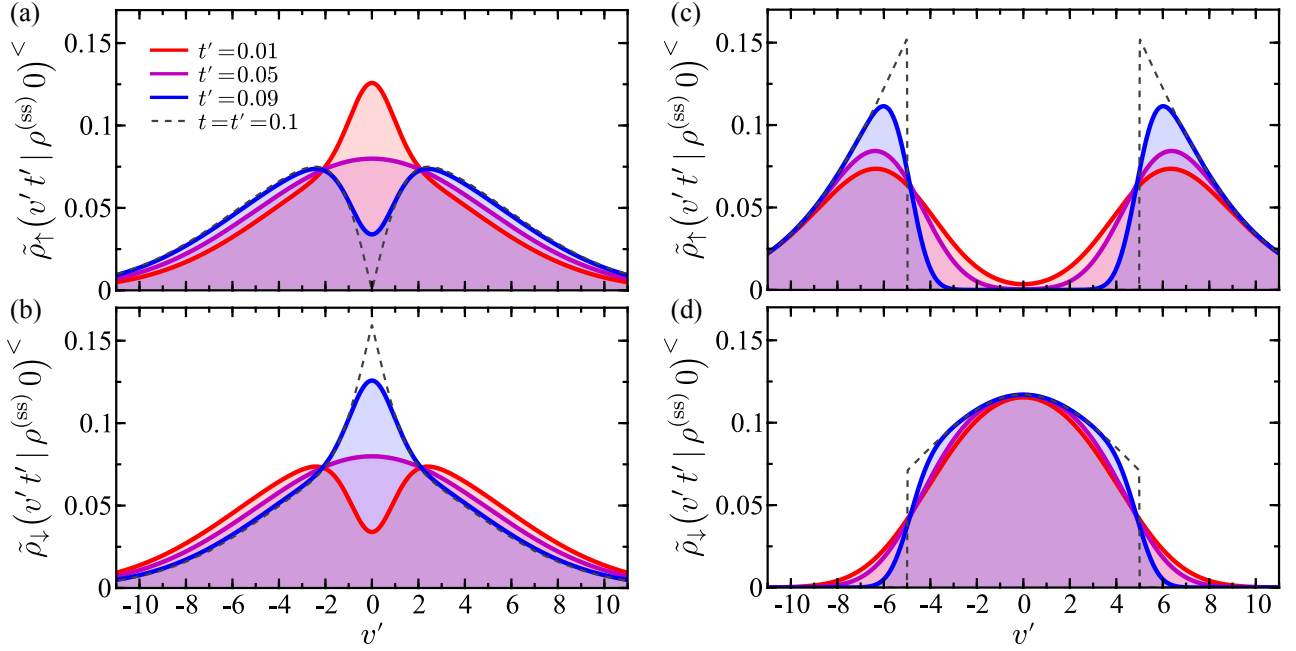}
\caption{\label{fig:probdenstwotime}
Two-time once-restricted upside and downside probability density for $\rho_0 = \rho^{(\text{ss})}$ with (a)-(b) $E^\ddag = E(0)$ and 
(c)-(d) $E^\ddag = \left\langle E\right\rangle$.
In all panels, the restricted densities are evaluated at different times $t'<t$ marked in the legend of (a)
and the upside/downside constraint is imposed at $t = 0.1$.
The dashed curves correspond to the respective densities at the $t' = t$ limit (equivalent densities are shown in Fig.~\ref{fig:probdens}).
}
\end{figure*}

In the case of $E^\ddag = E(0)$  and $\rho_0 = \rho^{(\text{ss})}$ there are no known closed-form expressions for 
Eqs.~(\ref{eq:densityupbackwardsgen}) and (\ref{eq:densitydownbackwardsgen}),
however, the restricted probability densities can be written using series representations:\cite{note2,Ng1969,Divsalar1998,Fayed2014,Fayed2015}
\begin{widetext}
\begin{align}
\label{eq:densityupbackwardsgenss}
 \nonumber \tilde{\rho}_\uparrow & \Big(v'\,t'<t\,\big|\,E(t)>E(0), \rho^{(\text{ss})}\,0 \Big)^< =  \\
\nonumber &\rho^{(\text{ss})}(v') \Bigg(2 - \frac{2}{\pi} 
\Bigg[\sum^\infty_{k = 0} \frac{(-1)^k a_+^{2 k+1}}{(2 k+1)} 
{_1F_1}\left(k+1,\tfrac{1}{2};-b_+^2\right)+\frac{(-1)^k \Gamma\big(k+\frac{1}{2}\big)a_+^{2 k}b_+}{\Gamma(k+1)} {_1F_1}\left(k+\tfrac{1}{2},\tfrac{3}{2};-b_+^2\right)\\
\nonumber& -\frac{(-1)^k a_-^{2 k+1}}{(2 k+1)} 
{_1F_1}\left(k+1,\tfrac{1}{2};-b_-^2\right)
-\frac{(-1)^k \Gamma\big(k+\frac{1}{2}\big)a_-^{2 k}b_-}{\Gamma(k+1)} {_1F_1}\left(k+\tfrac{1}{2},\tfrac{3}{2};-b_-^2\right)\\
\nonumber&  -\frac{(-1)^k a_+^{2 k+1}}{(2 k+1)} 
\mathcal{G}\big(k+1,\alpha^2\big)\,
{_1F_1}\left(k+1,\tfrac{1}{2};-b_+^2\right) \\
& \nonumber -\frac{\alpha (-1)^k \Gamma\big(k+\frac{1}{2}\big)a_+^{2 k}b_+}{|\alpha|\Gamma(k+1)}\mathcal{G}\big(k+\tfrac{1}{2},\alpha^2\big)\, {_1F_1}\left(k+\tfrac{1}{2},\tfrac{3}{2};-b_+^2\right)\\
\nonumber& +\frac{(-1)^k a_-^{2 k+1}}{(2 k+1)} 
\mathcal{G}\big(k+1,\alpha^2\big)\,
{_1F_1}\left(k+1,\tfrac{1}{2};-b_-^2\right) \\
& \nonumber +\frac{\alpha(-1)^k \Gamma\big(k+\frac{1}{2}\big)a_-^{2 k}b_-}{|\alpha|\Gamma(k+1)}
\mathcal{G}\big(k+\tfrac{1}{2},\alpha^2\big) \,{_1F_1}\left(k+\tfrac{1}{2},\tfrac{3}{2};-b_-^2\right)
\Bigg] \\
& -\erf\left[b_+\Big/\sqrt{a_+^2+1} \right] +\erf\left[b_-\Big/\sqrt{a_-^2+1} \right] \Bigg)
\end{align}
\end{widetext}
\begin{widetext}
\begin{align}
\label{eq:densitydownbackwardsgenss}
\nonumber  \tilde{\rho}_\downarrow & \Big(v'\,t'<t\,\big|\,E(t)<E(0), \rho^{(\text{ss})}\,0 \Big)^< =  \\
\nonumber &\rho^{(\text{ss})}(v') \Bigg(\frac{2}{\pi} 
\Bigg[\sum^\infty_{k = 0} \frac{(-1)^k a_+^{2 k+1}}{(2 k+1)} 
{_1F_1}\left(k+1,\tfrac{1}{2};-b_+^2\right)+\frac{(-1)^k \Gamma\big(k+\frac{1}{2}\big)a_+^{2 k}b_+}{\Gamma(k+1)} {_1F_1}\left(k+\tfrac{1}{2},\tfrac{3}{2};-b_+^2\right)\\
\nonumber& -\frac{(-1)^k a_-^{2 k+1}}{(2 k+1)} 
{_1F_1}\left(k+1,\tfrac{1}{2};-b_-^2\right)
-\frac{(-1)^k \Gamma\big(k+\frac{1}{2}\big)a_-^{2 k}b_-}{\Gamma(k+1)} {_1F_1}\left(k+\tfrac{1}{2},\tfrac{3}{2};-b_-^2\right)\\
\nonumber&  -\frac{(-1)^k a_+^{2 k+1}}{(2 k+1)} 
\mathcal{G}\big(k+1,\alpha^2\big)\,
{_1F_1}\left(k+1,\tfrac{1}{2};-b_+^2\right) \\
& \nonumber -\frac{\alpha (-1)^k \Gamma\big(k+\frac{1}{2}\big)a_+^{2 k}b_+}{|\alpha|\Gamma(k+1)}\mathcal{G}\big(k+\tfrac{1}{2},\alpha^2\big)\, {_1F_1}\left(k+\tfrac{1}{2},\tfrac{3}{2};-b_+^2\right)\\
\nonumber& +\frac{(-1)^k a_-^{2 k+1}}{(2 k+1)} 
\mathcal{G}\big(k+1,\alpha^2\big)\,
{_1F_1}\left(k+1,\tfrac{1}{2};-b_-^2\right)\\
& \nonumber +\frac{\alpha(-1)^k \Gamma\big(k+\frac{1}{2}\big)a_-^{2 k}b_-}{|\alpha|\Gamma(k+1)}
\mathcal{G}\big(k+\tfrac{1}{2},\alpha^2\big) \,{_1F_1}\left(k+\tfrac{1}{2},\tfrac{3}{2};-b_-^2\right)
\Bigg] \\
& +\erf\left[b_+\Big/\sqrt{a_+^2+1} \right] -\erf\left[b_-\Big/\sqrt{a_-^2+1} \right] \Bigg)
\end{align}
\end{widetext}
where
\begin{equation}
\mathcal{G}(a,z)  = \int_0^z v^{a-1} e^{-v} dv \Bigg/\int_0^\infty v^{a-1} e^{-v} dv,
\end{equation} 
is the normalized lower incomplete gamma function,
\begin{equation}
{_1F_1}\left(a,b;z\right) = \sum^\infty_{v = 0}\frac{(a)_v}{v!(b)_v}z^v,
\end{equation}
is a confluent hypergeometric function of the first kind, with 
\begin{equation}
(z)_v = z(z+1)(z+2) \cdots (z+v-1),
\end{equation} being the Pochhammer symbol, and
\begin{equation}
\begin{aligned}
a_\pm(t,t') &= \pm \sqrt{\frac{\sigma^2_v(t')}{\sigma^2_v(t-t')}}, \\
b_\pm(t,t',v') &= \left(e^{\gamma t'} \pm e^{\gamma (t-t')} \right) \sqrt{\left(\frac{1}{2 e^{2 \gamma t}\sigma^2_v(t-t')}\right)}v', \\
\alpha(t',v') &= -\sqrt{\frac{1}{2 e^{2 \gamma t'} \sigma^2_v(t')}}v'.
\end{aligned}
\end{equation}
The explicit dependence of $a_\pm$, $b_\pm$, and $\alpha$ on $t$, $t'$, and $v'$ has been suppressed in Eqs.~(\ref{eq:densityupbackwardsgenss}) and (\ref{eq:densitydownbackwardsgenss}) for notational convenience, but it should be understood that these quantities are functions.
In practice, we found that the computational time required to partially evaluate the infinite sums 
to the point of convergence was significant, and that application of quadrature methods 
to approximate the integrals in Eqs.~(\ref{eq:densityupbackwardsgen}) and (\ref{eq:densitydownbackwardsgen}) was a more efficient method.

As shown in Figs.~\ref{fig:probdenstwotime}(a)-(b),
for a system prepared with the steady-state distribution $\rho_0 = \rho^{(\text{ss})}$,
time-symmetry is observed in the restricted densities,
in that, at the midpoint $t' = t/2$ of the time interval $[0,t]$:
$\tilde{\rho}_\uparrow \equiv \tilde{\rho}_\downarrow$,
which implies that the upside and downside moments of $v'$ are 
equivalent
at this point.
Also note that for $t' \ll t$ the form of the upside/downside density takes the
shape of the opposite downside/upside density at $t' = t$ (cf. with the shapes in Figs.~\ref{fig:probdens}(a)-(b)).
In the limit $t' \to t$ the shape of the restricted probability densities
begins to morph smoothly into the shape of the densities 
for $t'= t$ (shown as a dashed curves)
which are equivalent to the
corresponding densities shown Figs.~\ref{fig:probdens}(a)-(b).


For threshold $E^\ddag = \left\langle E\right\rangle$ and initial distribution $\rho_0 = \rho^{(\text{ss})}$, the two-time restricted densities are:
\begin{align}
\label{eq:densityupbackwardsgenrho0Eavg}
\tilde{\rho}_\uparrow \Big(v'\,t'<t\,\big|\,\delta E^+, \rho^{(\text{ss})} \,0 \Big)^< &=  \rho^{(\text{ss})}(v')\Bigg[ \frac{\displaystyle 1- \Phi\big( \sqrt{\langle v^2 \rangle},v'; t-t'\big)+\Phi\big(- \sqrt{\langle v^2 \rangle},v'; t-t'\big)}
{ \displaystyle \erfc \left(\sqrt{1/2}\right)}\Bigg], \\
\label{eq:densitydownbackwardsgenrho0Eavg}
 \tilde{\rho}_\downarrow \Big(v'\,t'<t\,\big|\,\delta E^-, \rho^{(\text{ss})} \,0 \Big)^< &=  \rho^{(\text{ss})}(v') \Bigg[ \frac{\displaystyle \Phi\big( \sqrt{\langle v^2 \rangle} ,v'; t-t'\big)-\Phi\big(- \sqrt{\langle v^2 \rangle},v'; t-t'\big)}
{ \displaystyle \erf \left(\sqrt{1/2}\right)}\Bigg],
\end{align}
which are shown in Figs.~\ref{fig:probdenstwotime}(c)-(d)
as functions of $t'<t$.
Recall that, for this specific case, at $t' = t$ the shape of the restricted densities are independent of $t$ 
(see Fig.~\ref{fig:probdens}(c)-(d)),
which implies the distribution is stationary at the time where the
upside/downside constraint is imposed.
However, as shown in Figs.~\ref{fig:probdenstwotime}(c)-(d), for $t'<t$ the restricted densities are time-dependent.
Similar trends to the $E^\ddag  = E(0)$ case are observed in the time-evolution of $\tilde{\rho}_\uparrow$ and $\tilde{\rho}_\downarrow$ 
when using $\left\langle E\right\rangle$ as the threshold,
namely, the upside density takes a bimodal shape while the 
downside density takes a unimodal shape
and the densities are even functions of $v'$ for all $t'$.
In contrast with these geometrical similarities, prominent temporal differences arise in the evolution of the densities when using the different thresholds. 
Specifically, for threshold  $\left\langle E\right\rangle$ the 
densities are not time-symmetric, which differs from previous observations for threshold $E(0)$.
As $t' \to t$, the respective densities approach the functional forms given by Eqs.~(\ref{eq:transprobnegupdistthresfluc}) 
and (\ref{eq:transprobnegdowndistthresfluc}), which are shown as dashed curves, and
are equivalent to the stationary densities in Fig.~\ref{fig:probdens}(c)-(d).

\subsection{Two-time Restricted Moments: Velocity}

The two-time restricted densities can be applied to construct the velocity and energy moments of $v'(t'<t)$
given that the process is upside/downside at $t$.
The two-time once-restricted $k$th moments of $v'$ are 
\begin{align}
\nonumber \Big\langle v'^k\big(t'\big|\,E(t)>E^\ddag,\, \rho_0 \,0\big)\Big\rangle^{<}_\uparrow &\equiv \big\langle v^k(t')\big\rangle_\uparrow^< \\
& = \frac{ \displaystyle \int_{\mathbb{R}^3}\!\!\! v'^k \rho_0(\bar{v}) \rho\big(v\, t  | \, v'\,  t'\big) 
\rho\big(v'\,  t'\, | \, \bar{v}\, 0\big)\Theta\big(E(v)-E^\ddag\big)\,d\bar{v}\,dv'dv}{\displaystyle \int_{\mathbb{R}^3}\!\!\! \rho_0(\bar{v}) \rho\big(v\, t  | \, v'\,  t'\big) \rho\big(v'\,  t'\, | \, \bar{v}\, 0\big)\Theta\big(E(v)-E^\ddag\big)\,d\bar{v}\,dv'dv},\\[1ex]
 \nonumber \Big\langle  v'^k \big(t'\big|\,E(t)<E^\ddag,\, \rho_0\,0 \big)\Big\rangle^{<}_\downarrow &\equiv \big\langle v^k(t')\big\rangle_\downarrow^< \\
& =  \frac{\displaystyle \int_{\mathbb{R}^3}\!\!\! v'^k \rho_0(\bar{v}) \rho\big(v\, t  | \, v'\,  t'\big) \rho\big(v'\,  t'\, | \, \bar{v}\, 0\big)
\Theta\big(E^\ddag-E(v)\big)\,d\bar{v}\,dv'dv}{\displaystyle \int_{\mathbb{R}^3}\!\!\! \rho_0(\bar{v}) \rho\big(v\, t  | \, v'\,  t'\big) \rho\big(v'\,  t'\, | \, \bar{v}\, 0\big)\Theta\big(E^\ddag-E(v)\big)\,d\bar{v}\,dv'dv}.
\end{align}
In the limit $t' \to t$ the ``one-time" forms derived in Sec.~\ref{sec:restvel}:
\begin{equation}
\begin{aligned}
\label{eq:relationtandtprime}
 \Big\langle  v'^k\big(t'=t\,\big|\,E(t)>E^\ddag, \rho_0 \,0 \big)\Big\rangle_{\uparrow} &= \big\langle  v^k(t )\big\rangle_{\uparrow}, \\[0ex]
 \Big\langle  v'^k\big(t'=t\,\big|\,E(t)<E^\ddag, \rho_0 \,0 \big)\Big\rangle_{\uparrow} &= \big\langle  v^k(t )\big\rangle_{\downarrow}, 
\end{aligned}
\end{equation}
are recovered.


\begin{figure}[]
\includegraphics[width = 8.5cm,clip]{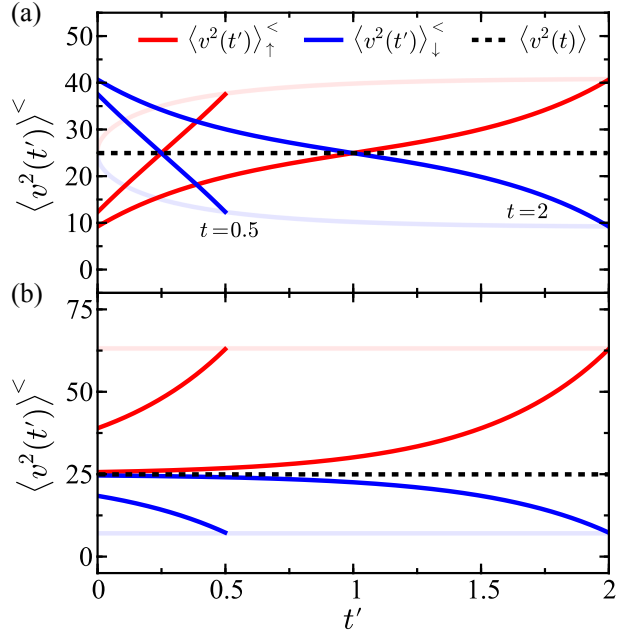}
\caption{\label{fig:velsq_ss_twotime}
Two-time once-restricted velocity moment $\langle v^2(t') \rangle^<$ 
as a function of $t'$ for $t = 0.5$ and $t = 2$
and energy thresholds (a) $E(0)$
and (b) $\left\langle E\right\rangle$.
The light transparent curves (red; upside and blue; downside) are the $\langle v^2(t' = t) \rangle^<$ results.
In both panels $\rho_0 = \rho^{(\text{ss})}$.
The dashed black line corresponds to the unrestricted moment $\langle v^2(t) \rangle$.
}
\end{figure}

\textit{$E(t)$ compared to $E(0)$}---
For a system with $\rho_0 = \rho^{(\text{ss})}$
and $E^\ddag = E(0)$
the first two-time restricted velocity moments vanish:
\begin{equation}
\label{eq:vtss}
\Big\langle v'\big(t'\big|\, \rho^{(\text{ss})} \,0\big) \Big\rangle^{<}_\uparrow = \Big\langle v'\big(t'\big|\, \rho^{(\text{ss})} \,0 \big)\Big\rangle^{<}_\downarrow  =0.
\end{equation}
This arises from symmetry with respect to $v'$ in the corresponding restricted two-time probability densities
shown in Figs.~\ref{fig:probdenstwotime}(c)-(d).
The second moments:
\begin{align}
\label{eq:vsq2timeup}
\Big\langle v'^2\big(t' \big|\,E(t)>E(0), \rho^{(\text{ss})}\, 0 \big)\Big\rangle^{<}_\uparrow &= \frac{k_\text{B} T}{m} \Bigg[1-\frac{2}{\pi}\Bigg( \frac{e^{-2 \gamma t'}-e^{-2\gamma(t-t')}}{G(t)}\Bigg)\Bigg],\\
\label{eq:vsq2timedown}
 \Big\langle v'^2\big(t'  \big|\,E(t)<E(0), \rho^{(\text{ss})}\, 0 \big)\Big\rangle^{<}_\downarrow &= \frac{k_\text{B} T}{m} \Bigg[1+\frac{2}{\pi} \Bigg(\frac{e^{-2 \gamma t'}-e^{-2\gamma(t-t')}}{G(t)}\Bigg)\Bigg],
\end{align}
with $G(t)$ taken from Eq.~(\ref{eq:G}),
are shown in Fig.~\ref{fig:velsq_ss_twotime}(a)
as functions of $t'$.
At $t' = t/2$ the second moments are equivalent, and are equal to the stationary unrestricted moment:
$\langle v^2(t/2)\rangle_\uparrow^< = \langle v^2(t/2)\rangle_\downarrow^< = \langle v^2 \rangle$.
This is a direct consequence of the equivalence of the upside and downside densities at this timepoint as shown in Figs.~\ref{fig:probdenstwotime}(c)-(d).
In the $t' \to t$ limit the relations in Eq.~(\ref{eq:relationtandtprime}) are satisfied and the respective two-time moments are
equivalent to $\langle v^2(t)\rangle_\uparrow$ and $\langle v^2(t)\rangle_\downarrow$ which are given by Eqs.~(\ref{eq:vsqssup}) and (\ref{eq:vsqssdown}).
In the opposite $t' \to 0$ limit, time-symmetry is observed in that 
$\langle v^2(0)\rangle_\uparrow^< = \langle v^2(t)\rangle_\downarrow^<$ and $\langle v^2(0)\rangle_\downarrow^< = \langle v^2(t)\rangle_\uparrow^<$.
This implies that the upside/downside second velocity moment at $t' = 0$ is equal to the downside/upside second moment at $t' = t$. 
The origin of this time-symmetry from the underlying respective restricted densities can be seen in Figs.~\ref{fig:probdenstwotime}(c)-(d).

\textit{$E(t)$ relative to  $\left\langle E\right\rangle$}---
The two-time once-restricted first velocity moments for energy fluctuations from the average under steady-state conditions are
\begin{equation}
\label{eq:vtssavg}
\Big\langle v'\big(t'\big|\,\delta E^+, \rho^{(\text{ss})}\, 0 \big)\Big\rangle^{<}_\uparrow = \Big\langle v'\big(t'\big|\,\delta E^-, \rho^{(\text{ss})}\, 0 \big) \Big\rangle^{<}_\downarrow  =0,
\end{equation}
and the corresponding second moments are
\begin{align}
\label{eq:vtpupfluc}
\Big\langle v'^2\big(t'\big|\,\delta E^+, \rho^{(\text{ss})}\, 0 \big)\Big\rangle^{<}_\uparrow &= \frac{k_\text{B} T} {m}\left[1+\sqrt{\frac{2}{\pi e}} \left(\frac{ e^{-2 \gamma (t-t')}}{\erfc{(\sqrt{1/2})}}\right)\right],
\\[1ex]
\label{eq:vtpdownfluc}
\Big\langle v'^2\big(t'\big|\,\delta E^-, \rho^{(\text{ss})}\, 0 \big)\Big\rangle^{<}_\downarrow  &= \frac{k_\text{B} T} {m}\left[1-\sqrt{\frac{2}{\pi e}} \left(\frac{ e^{-2 \gamma (t-t')}}{\erf{(\sqrt{1/2})}}\right)\right].
\end{align}
Figure~\ref{fig:velsq_ss_twotime}(b)
illustrates the dependence of the second restricted moments on $t'$ for various values of $t$.
As in previous cases, in the limit $t' \to t$ the relations between the two-time moments are given by Eq.~(\ref{eq:relationtandtprime}).
In the $t \to 0$ limit,
$\langle v^2(0)\rangle_\uparrow^< \to \langle v^2\rangle_\uparrow$ and 
$\langle v^2(0)\rangle_\downarrow^< \to \langle v^2\rangle_\downarrow$
which are given by Eqs.~(\ref{eq:vsqssupfluc}) and (\ref{eq:vsqssdownfluc}), respectively. 
As $t$ is increased and $t' \to 0$, the two-time restricted second moments approach  
the stationary unrestricted value $\langle v^2\rangle$.
This implies that for large $t$ (in relation to $\gamma$) 
the statistical properties for $t' \ll t$ are given by the unrestricted properties.

\subsection{Two-time Restricted Moments: Energy Change and Energy Fluctuation}

\textit{$E(t)$ compared to $E(0)$}---
The general expressions for the change in energy $\Delta E = E-E(0)$ during upside and downside processes 
for threshold $E^\ddag = E(0)$ and initial distribution $\rho_0$ are
\begin{align}
\label{eq:DeltaEup}
\nonumber \Big\langle \Delta E\big(t\,\big|\,E(t)>E(0),  \rho_0 \,0\big)\Big\rangle_\uparrow   & =  \Big\langle  E\big(t\,\big|\,E(t)>E(0),  \rho_0 \,0\big)\Big\rangle_\uparrow  \\
&\quad -  \Big\langle  E\big(t' = 0\,\big|\,E(t)>E(0),  \rho_0 \,0\big)\Big\rangle^<_\uparrow , \\[0ex]
\label{eq:DeltaEdown}
 \nonumber \Big\langle \Delta E\big(t\,\big|\,E(t)<E(0),  \rho_0 \,0\big)\Big\rangle_\downarrow   &= 
  \Big\langle  E\big(t\,\big|\,E(t)<E(0), \rho_0 \,0\big)\Big\rangle_\downarrow  \\
&\quad -\Big\langle  E\big(t' = 0\,\big|\,E(t)>E(0),  \rho_0 \,0\big)\Big\rangle^<_\downarrow,
\end{align}
where the first term on the R.H.S. of each equation is the energy of the process at time $t$ (which is the time where the upside/downside constraint is imposed) and the second term on the R.H.S. is the energy of the process at time $t'=0$ given that it is upside/downside at time $t$. 
Recall that the ``$<$'' superscript denotes that the upside/downside constraint is imposed at time $t$ while observable is evaluated at $t'<t$.



In the case of a system that is initially characterized by the distribution $\rho^{(\text{ss})}$,
the restricted expectation values for the system energy at time $t$ are
\begin{align}
  &\Big\langle  E\big(t\,\big|\,E(t)>E(0), \rho^{(\text{ss})} \,0\big)\Big\rangle_\uparrow    = \frac{k_\text{B} T} { 2} \bigg[1+\frac{2}{\pi} G(t) \bigg],\\[1ex]
	&\Big\langle  E\big(t\,\big|\,E(t)<E(0), \rho^{(\text{ss})} \,0\big)\Big\rangle_\downarrow   = \frac{k_\text{B} T} { 2} \bigg[1-\frac{2}{\pi} G(t) \bigg],
\end{align}
and expectation values of the energy of the system at $t' = 0$ given that it is upside/downside at $t$ are
\begin{align}
 &\Big\langle  E\big(t' = 0\,\big|\,E(t)>E(0), \rho^{(\text{ss})} \,0\big)\Big\rangle^<_\uparrow    = \frac{k_\text{B} T} { 2} \bigg[1-\frac{2}{\pi} G(t) \bigg],\\[1ex]
 &\Big\langle  E\big(t' = 0\,\big|\,E(t)<E(0), \rho^{(\text{ss})} \,0\big)\Big\rangle^<_\downarrow   = \frac{k_\text{B} T} { 2} \bigg[1+\frac{2}{\pi} G(t) \bigg].
\end{align}
Note that application of this threshold and initial distribution give rise to the peculiar property that the expected energy of an upside/downside process at time $t$ is the same as the expected energy of 
the conjugate downside/upside process at  $t' = 0$. 
Therefore, the restricted energy changes are
\begin{align}
 &\Big\langle  \Delta E\big(t\,\big|\,E(t)>E(0), \rho^{(\text{ss})} \,0\big)\Big\rangle_\uparrow    = \frac{2 k_\text{B} T }{ \pi }G(t),\\[1ex] 
 &\Big\langle  \Delta E\big(t\,\big|\,E(t)<E(0), \rho^{(\text{ss})} \,0\big)\Big\rangle_\downarrow   = -\frac{2 k_\text{B} T  }{\pi } G(t). 
\end{align}

\textit{$E(t)$ relative to  $\left\langle E\right\rangle$}---
In the case of energy threshold $E^\ddag = \langle E \rangle$,
an energy fluctuation,
\begin{equation}
\delta E \equiv E(t)- \langle E \rangle,
\end{equation}
is defined as a deviation from the average energy. 
The general expressions for the expectation value of the magnitude of restricted fluctuations given that the system is characterized by 
distribution $\rho_0$ are 
\begin{align}
\label{eq:DeltaEupavg}
 \Big\langle \delta E\big(t\,\big|\,\delta E^+,  \rho_0 \,0\big)\Big\rangle_\uparrow    &=  \Big\langle  E\big(t\,\big|\,\delta E^+,  \rho_0 \,0\big)\Big\rangle_\uparrow -  \langle E \rangle, \\[0ex]
\label{eq:DeltaEdownavg}
\Big\langle \delta E\big(t\,\big|\,\delta E^-,  \rho_0 \,0\big)\Big\rangle_\downarrow   &=  \Big\langle  E\big(t\,\big|\,\delta E^-, \rho_0 \,0\big)\Big\rangle_\downarrow - \langle E \rangle.
\end{align}
In the specific case of $\rho_0 = \rho^{(\text{ss})}$, application of Eqs.~(\ref{eq:DeltaEupavg}) and (\ref{eq:DeltaEdownavg})
coupled with Eqs.~(\ref{eq:avgenergy}), (\ref{eq:vtpupfluc}), and (\ref{eq:vtpdownfluc})
yields:
\begin{align}
 &\Big\langle  \delta E\big(t\,\big|\,\delta E^+, \rho^{(\text{ss})} \,0\big)\Big\rangle_\uparrow    = \sqrt{\frac{1}{2 \pi e}}\left(\frac{k_\text{B} T }{\erfc{(\sqrt{1/2})}}\right),\\[0ex] 
 &\Big\langle  \delta E\big(t\,\big|\,\delta E^-, \rho^{(\text{ss})} \,0\big)\Big\rangle_\downarrow   = -\sqrt{\frac{1}{2 \pi e}}\left(\frac{k_\text{B} T }{ \erf{(\sqrt{1/2})}}\right),
\end{align}
which are time-independent.

The expectation values of the energy change for a process that is
characterized by distribution $\rho_0$
and a positive or negative energy fluctuation at time $t$ are:
\begin{align}
\label{eq:DeltaEupavgfluc}
 \nonumber\Big\langle  \Delta E\big(t\,\big|\,\delta E^+,  \rho_0 \,0 \big)\Big\rangle_\uparrow  &= 
\Big\langle \delta E\big(t\,\big|\,\delta E^+,  \rho_0 \,0\big)\Big\rangle_\uparrow  \\  & \quad -
 \Big\langle  E\big(t' = 0\,\big|\,\delta E^+, \rho_0 \,0\big)\Big\rangle_\uparrow +  \langle E \rangle, \\[1ex]
\label{eq:DeltaEdownavgfluc}
\nonumber\Big\langle  \Delta E\big(t\,\big|\,\delta E^-,  \rho_0 \,0 \big)\Big\rangle_\downarrow   &= 
\Big\langle \delta E\big(t\,\big|\,\delta E^-,  \rho_0 \,0\big)\Big\rangle_\downarrow  \\ &\quad  - 
\Big\langle  E\big(t' = 0\,\big|\,\delta E^-, \rho_0 \,0\big)\Big\rangle_\downarrow + \langle E \rangle.
\end{align}
When applying this threshold, $\Delta E$ can be positive or negative for an upside trajectory 
and likewise for a downside trajectory.
This is because, in this case, the upside/downside criterion is that the process energy be above the threshold at time $t$,
not that the process energy has increased or decreased with respect to its initial value.
In the special case $\rho_0 = \rho^{(\text{ss})}$ we get
\begin{align}
\label{eq:DeltaEupavgflucss}
\Big\langle  \Delta E\big(t\,\big|\,\delta E^+,  \rho^{(\text{ss})} \,0 \big)\Big\rangle_\uparrow   &=  \sqrt{\frac{1}{2 \pi e}}\left(\frac{k_\text{B} T }{\erfc{(\sqrt{1/2})}}\right)\Big(1- e^{-2 \gamma t}\Big), \\[1ex]
\label{eq:DeltaEdownavgflucss}
\Big\langle  \Delta E\big(t\,\big|\,\delta E^-,  \rho^{(\text{ss})} \,0 \big)\Big\rangle_\downarrow   &=  -\sqrt{\frac{1}{2 \pi e}}\left(\frac{k_\text{B} T }{\erf{(\sqrt{1/2})}}\right)\Big(1- e^{-2 \gamma t}\Big).
\end{align}

\subsection{Restricted Velocity Correlation Functions}

The two-time once-restricted velocity correlation functions 
$\left\langle v'(t')v(t)\right\rangle_\uparrow^<$ and $\left\langle v'(t')v(t)\right\rangle_\downarrow^<$ 
relate the velocity at times $t'<t$ and $t$ (denoted by $v'$ and $v$) given that the process
is upside/downside at $t$.
These functions, which quantify the timescale of relaxation or activation,
can be constructed using similar methods to those applied previously to obtain the two-time restricted moments.
The general expressions for these correlations are:
\begin{align}
\label{eq:gencorrup}
\nonumber \big\langle v'(t') v(t)|\,E(t)>E^\ddag,\, \rho_0 \,0\big\rangle^{<}_\uparrow &\equiv \big\langle v'(t')v(t)\big\rangle_\uparrow^< \\
&= \frac{ \displaystyle \int_{\mathbb{R}^3}\!\!\! v'v \rho_0(\bar{v}) \rho\big(v\, t  | \, v'\,  t'\big) 
\rho\big(v'\,  t'\, | \, \bar{v}\, 0\big)\Theta\big(E(v)-E^\ddag\big)\,d\bar{v}\,dv'dv}{\displaystyle \int_{\mathbb{R}^3}\!\!\! \rho_0(\bar{v}) \rho\big(v\, t  | \, v'\,  t'\big) \rho\big(v'\,  t'\, | \, \bar{v}\, 0\big)\Theta\big(E(v)-E^\ddag\big)\,d\bar{v}\,dv'dv},\\[1ex]
\label{eq:gencorrdown}
 \nonumber \big\langle  v'(t')v(t)|\,E(t)<E^\ddag,\, \rho_0\,0\big\rangle^{<}_\downarrow &\equiv \big\langle v'(t')v(t)\big\rangle_\downarrow^< \\
&= \frac{\displaystyle \int_{\mathbb{R}^3}\!\!\! v'v \rho_0(\bar{v}) \rho\big(v\, t  | \, v'\,  t'\big) \rho\big(v'\,  t'\, | \, \bar{v}\, 0\big)
\Theta\big(E^\ddag-E(v)\big)\,d\bar{v}\,dv'dv}{\displaystyle \int_{\mathbb{R}^3}\!\!\! \rho_0(\bar{v}) \rho\big(v\, t  | \, v'\,  t'\big) \rho\big(v'\,  t'\, | \, \bar{v}\, 0\big)\Theta\big(E^\ddag-E(v)\big)\,d\bar{v}\,dv'dv}.
\end{align}

\begin{figure}[]
\includegraphics[width = 8.5cm,clip]{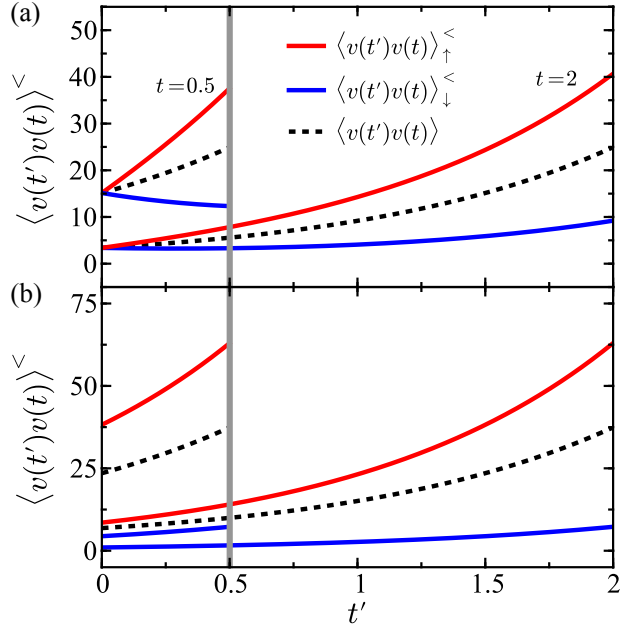}
\caption{\label{fig:velcorr}
Velocity correlation as a function of $t'$ for energy threshold (a) $E(0)$ and 
(b) $\left\langle E \right\rangle$.
In both panels $\rho_0 = \rho^{(\text{ss})}$.
Sets of curves are shown for $t = 0.5$ (denoted by a thick vertical line) and $t = 2$.
In each panel the unrestricted velocity correlation function for each value of $t$ is shown as a dashed curve.
}
\end{figure}

\textit{$E(t)$ compared to $E(0)$}---
For initial distribution $\rho_0 = \rho^{(\text{ss})}$, evaluating Eqs.~(\ref{eq:gencorrup}) and (\ref{eq:gencorrdown}) yields
\begin{align}
\label{eq:vtpupv0crossss}
  \Big\langle v'(t') v(t) \big|\,E(t)>E(0), \rho^{(\text{ss})}\, 0 \Big\rangle^{<}_\uparrow &= \frac{k_\text{B} T}{m} \Bigg( e^{-\gamma(t-t')} + \frac{4}{\pi} \frac{\sinh[\gamma t']}{G(t)}e^{-\gamma t}\Bigg), 
\\[1ex]
\label{eq:vtpdownv0crossss}
 \Big\langle v'(t')v(t)\big|\,E(t)<E(0), \rho^{(\text{ss})}\, 0 \Big\rangle^{<}_\downarrow &=  \frac{k_\text{B} T}{m} \Bigg( e^{-\gamma(t-t')} - \frac{4}{\pi} \frac{\sinh[\gamma t']}{G(t)}e^{-\gamma t}\Bigg),
\end{align}
where $G(t)$ is given by Eq.~(\ref{eq:G}).
These correlation functions are shown in Fig.~\ref{fig:velcorr}(a) over variation in $t'$.
In the $t' \to 0$ limit the unrestricted velocity correlation function 
and both restricted velocity correlation functions approach $\langle v^2 \rangle e^{-\gamma t}$.
This limiting behavior illustrates that the distributions of $v'$ and $v$ are equivalent in this limit.
In the opposite $t' \to t$ limit, all correlations recover the respective unrestricted/restricted expectation value of the velocity squared given by
Eqs.~(\ref{eq:vsqunrestricted}), (\ref{eq:vsqssup}), and (\ref{eq:vsqssdown}).

\textit{$E(t)$ relative to  $\left\langle E\right\rangle$}---
In the case that upside/downside events are defined by positive/negative energy fluctuations, the restricted velocity correlation functions for a system prepared under steady-state conditions are
\begin{align}
\label{eq:vtpupfluccross}
\Big\langle v'(t') v(t)\big|\,\delta E^+, \rho^{(\text{ss})}\, 0 \Big\rangle^{<}_\uparrow &= \big\langle v^2(t)\big\rangle_\uparrow e^{-\gamma (t-t')},
\\[1ex]
\label{eq:vtpdownfluccross}
\Big\langle v'(t') v(t)\big|\,\delta E^-, \rho^{(\text{ss})}\, 0 \Big\rangle^{<}_\downarrow &=  \big\langle v^2(t)\big\rangle_\downarrow e^{-\gamma (t-t')},
\end{align}
where the one-time restricted second moments are taken from
Eqs.~(\ref{eq:vsqssupfluc}) and (\ref{eq:vsqssdownfluc}).
The restricted correlation functions take 
a simple form that is analogous to the unrestricted velocity correlation function.
Figure~\ref{fig:velcorr}(b) illustrates these correlations 
as a function of $t'$ for several values of $t$. 
In the limit $t' \to 0$  the unrestricted correlation approaches $\langle v^2 \rangle e^{-\gamma t}$
and the upside and downside correlations approach, respectively,
$2.53 \times \langle v^2 \rangle e^{-\gamma t}$ and $0.291 \times \langle v^2 \rangle e^{-\gamma t}$.
Therefore, as $t \to \infty$ with finite $t'$ the velocities $v'(t')$ and $v(t)$ are uncorrelated
for both unrestricted and restricted transport, as expected.
In the $t' \to t$ limit the velocity correlations approach the respective expectation values of the squared velocity
given in the unrestricted case by Eq.~(\ref{eq:vsqunrestricted}) and in the restricted cases by 
Eqs.~(\ref{eq:vsqssupfluc}) and (\ref{eq:vsqssdownfluc}).

\section{\label{sec:conc}Conclusions}

We have presented a statistical analysis of
sets of Brownian trajectories 
that have been separated from the full ensemble
using the criterion that all the trajectories in each set are either above or below an energy threshold
as some given time.
This formalism has been applied to reveal transport properties that arise when treating, separately,
upside (energy activation) and downside (energy relaxation) events, and it has been shown that these properties
differ in both their steady-state form and temporal evolution from those obtained through analysis of the full ensemble.
Specifically, explicit forms for upside/downside velocity distributions, moments, and correlation functions 
have been derived for several pertinent energy thresholds and initial distributions
with particular importance in nonequilibrium statistical mechanics.

The focus of this article has been development of the upside/downside mathematical framework.
In subsequent articles, we apply this formalism to
the analysis of energy partitioning during upside and downside events under steady-state conditions 
for a particle that is coupled to multiple thermal 
baths characterized by different temperatures.
Particular focus is placed on upside/downside nonequilibrium thermal transport properties
and their relation to the traditional theoretical picture of heat conduction
in which fluctuations are treated as properties intrinsic to the full ensemble.
It has been shown that this analysis is pertinent for evaluating heat transport driven by activated chemical processes that take place 
in thermally heterogeneous environments. \cite{craven17e}

\section{Supplementary Material}

The Supplementary Material contains derivations of the restricted transition probabilities, distributions, moments, and correlation functions
for cases with: $E^\ddag = E(0)$ and $\rho_0 = \rho^{(\text{ss})}$, $E^\ddag = E(0)$ and $\rho_0 = \delta(v-v_0)$,
$E^\ddag = \left\langle E\right\rangle$ and $\rho_0 = \rho^{(\text{ss})}$.

\section{Acknowledgments}

The research of AN is supported by the Israel-U.S. Binational Science Foundation, 
the German Research Foundation (DFG TH 820/11-1), 
the U.S. National Science Foundation (Grant No. CHE1665291),
and the University of Pennsylvania.

\bibliography{c6_1_combined}

\begin{thebibliography}{57}
\expandafter\ifx\csname natexlab\endcsname\relax\def\natexlab#1{#1}\fi
\expandafter\ifx\csname bibnamefont\endcsname\relax
  \def\bibnamefont#1{#1}\fi
\expandafter\ifx\csname bibfnamefont\endcsname\relax
  \def\bibfnamefont#1{#1}\fi
\expandafter\ifx\csname citenamefont\endcsname\relax
  \def\citenamefont#1{#1}\fi
\expandafter\ifx\csname url\endcsname\relax
  \def\url#1{\texttt{#1}}\fi
\expandafter\ifx\csname urlprefix\endcsname\relax\def\urlprefix{URL }\fi
\providecommand{\bibinfo}[2]{#2}
\providecommand{\eprint}[2][]{\url{#2}}

\bibitem[{\citenamefont{Jarzynski}(1997)}]{Jarzynski1997}
\bibinfo{author}{\bibfnamefont{C.}~\bibnamefont{Jarzynski}},
  \bibinfo{journal}{Phys. Rev. Lett.} \textbf{\bibinfo{volume}{78}},
  \bibinfo{pages}{2690} (\bibinfo{year}{1997}),
  \eprint{doi:10.1103/PhysRevLett.78.2690}.

\bibitem[{\citenamefont{Evans et~al.}(1993)\citenamefont{Evans, Cohen, and
  Morriss}}]{Evans1993}
\bibinfo{author}{\bibfnamefont{D.~J.} \bibnamefont{Evans}},
  \bibinfo{author}{\bibfnamefont{E.~G.~D.} \bibnamefont{Cohen}},
  \bibnamefont{and} \bibinfo{author}{\bibfnamefont{G.~P.}
  \bibnamefont{Morriss}}, \bibinfo{journal}{Phys. Rev. Lett.}
  \textbf{\bibinfo{volume}{71}}, \bibinfo{pages}{2401} (\bibinfo{year}{1993}),
  \eprint{doi:10.1103/PhysRevLett.71.2401}.

\bibitem[{\citenamefont{Kurchan}(1998)}]{Kurchan1998}
\bibinfo{author}{\bibfnamefont{J.}~\bibnamefont{Kurchan}}, \bibinfo{journal}{J.
  Phys. A} \textbf{\bibinfo{volume}{31}}, \bibinfo{pages}{3719}
  (\bibinfo{year}{1998}),
  \eprint{http://stacks.iop.org/0305-4470/31/i=16/a=003}.

\bibitem[{\citenamefont{Crooks}(2000)}]{Crooks2000}
\bibinfo{author}{\bibfnamefont{G.~E.} \bibnamefont{Crooks}},
  \bibinfo{journal}{Phys. Rev. E} \textbf{\bibinfo{volume}{61}},
  \bibinfo{pages}{2361} (\bibinfo{year}{2000}),
  \eprint{doi:10.1103/PhysRevE.61.2361}.

\bibitem[{\citenamefont{Seifert}(2012)}]{Seifert2012}
\bibinfo{author}{\bibfnamefont{U.}~\bibnamefont{Seifert}},
  \bibinfo{journal}{Rep. Prog. Phys.} \textbf{\bibinfo{volume}{75}},
  \bibinfo{pages}{126001} (\bibinfo{year}{2012}),
  \eprint{http://stacks.iop.org/0034-4885/75/i=12/a=126001}.

\bibitem[{\citenamefont{Onsager}(1931)}]{Onsager1931}
\bibinfo{author}{\bibfnamefont{L.}~\bibnamefont{Onsager}},
  \bibinfo{journal}{Phys. Rev.} \textbf{\bibinfo{volume}{37}},
  \bibinfo{pages}{405} (\bibinfo{year}{1931}),
  \eprint{doi:10.1103/PhysRev.37.405}.

\bibitem[{\citenamefont{Sekimoto}(1998)}]{Sekimoto1998}
\bibinfo{author}{\bibfnamefont{K.}~\bibnamefont{Sekimoto}},
  \bibinfo{journal}{Prog. Theor. Phys. Supp.} \textbf{\bibinfo{volume}{130}},
  \bibinfo{pages}{17} (\bibinfo{year}{1998}), \eprint{doi:10.1143/PTPS.130.17}.

\bibitem[{\citenamefont{Seifert}(2005)}]{Seifert2005}
\bibinfo{author}{\bibfnamefont{U.}~\bibnamefont{Seifert}},
  \bibinfo{journal}{Phys. Rev. Lett.} \textbf{\bibinfo{volume}{95}},
  \bibinfo{pages}{040602} (\bibinfo{year}{2005}),
  \eprint{doi:10.1103/PhysRevLett.95.040602}.

\bibitem[{\citenamefont{Van~den Broeck}(2013)}]{Van2013stochastic}
\bibinfo{author}{\bibfnamefont{C.}~\bibnamefont{Van~den Broeck}}, in
  \emph{\bibinfo{booktitle}{Physics of Complex Colloids}}
  (\bibinfo{publisher}{IOS Press}, \bibinfo{year}{2013}), vol.
  \bibinfo{volume}{184}, pp. \bibinfo{pages}{155--193}.

\bibitem[{\citenamefont{Lebowitz}(1959)}]{Lebowitz1959}
\bibinfo{author}{\bibfnamefont{J.~L.} \bibnamefont{Lebowitz}},
  \bibinfo{journal}{Phys. Rev.} \textbf{\bibinfo{volume}{114}},
  \bibinfo{pages}{1192} (\bibinfo{year}{1959}),
  \eprint{doi:10.1103/PhysRev.114.1192}.

\bibitem[{\citenamefont{Rieder et~al.}(1967)\citenamefont{Rieder, Lebowitz, and
  Lieb}}]{Lebowitz1967}
\bibinfo{author}{\bibfnamefont{Z.}~\bibnamefont{Rieder}},
  \bibinfo{author}{\bibfnamefont{J.~L.} \bibnamefont{Lebowitz}},
  \bibnamefont{and} \bibinfo{author}{\bibfnamefont{E.}~\bibnamefont{Lieb}},
  \bibinfo{journal}{J. Math. Phys.} \textbf{\bibinfo{volume}{8}},
  \bibinfo{pages}{1073} (\bibinfo{year}{1967}), \eprint{doi:10.1063/1.1705319}.

\bibitem[{\citenamefont{Casher and Lebowitz}(1971)}]{Lebowitz1971}
\bibinfo{author}{\bibfnamefont{A.}~\bibnamefont{Casher}} \bibnamefont{and}
  \bibinfo{author}{\bibfnamefont{J.~L.} \bibnamefont{Lebowitz}},
  \bibinfo{journal}{J. Math. Phys.} \textbf{\bibinfo{volume}{12}},
  \bibinfo{pages}{1701} (\bibinfo{year}{1971}), \eprint{doi:10.1063/1.1665794}.

\bibitem[{\citenamefont{Segal et~al.}(2003)\citenamefont{Segal, Nitzan, and
  H\"anggi}}]{Nitzan2003thermal}
\bibinfo{author}{\bibfnamefont{D.}~\bibnamefont{Segal}},
  \bibinfo{author}{\bibfnamefont{A.}~\bibnamefont{Nitzan}}, \bibnamefont{and}
  \bibinfo{author}{\bibfnamefont{P.}~\bibnamefont{H\"anggi}},
  \bibinfo{journal}{J. Chem. Phys.} \textbf{\bibinfo{volume}{119}},
  \bibinfo{pages}{6840} (\bibinfo{year}{2003}), \eprint{doi:10.1063/1.1603211}.

\bibitem[{\citenamefont{Segal and Nitzan}(2005)}]{Segal2005prl}
\bibinfo{author}{\bibfnamefont{D.}~\bibnamefont{Segal}} \bibnamefont{and}
  \bibinfo{author}{\bibfnamefont{A.}~\bibnamefont{Nitzan}},
  \bibinfo{journal}{Phys. Rev. Lett.} \textbf{\bibinfo{volume}{94}},
  \bibinfo{pages}{034301} (\bibinfo{year}{2005}),
  \eprint{doi:10.1103/PhysRevLett.94.034301}.

\bibitem[{\citenamefont{Dhar and Lebowitz}(2008)}]{Lebowitz2008}
\bibinfo{author}{\bibfnamefont{A.}~\bibnamefont{Dhar}} \bibnamefont{and}
  \bibinfo{author}{\bibfnamefont{J.~L.} \bibnamefont{Lebowitz}},
  \bibinfo{journal}{Phys. Rev. Lett.} \textbf{\bibinfo{volume}{100}},
  \bibinfo{pages}{134301} (\bibinfo{year}{2008}),
  \eprint{doi:10.1103/PhysRevLett.100.134301}.

\bibitem[{\citenamefont{Kannan et~al.}(2012)\citenamefont{Kannan, Dhar, and
  Lebowitz}}]{Lebowitz2012}
\bibinfo{author}{\bibfnamefont{V.}~\bibnamefont{Kannan}},
  \bibinfo{author}{\bibfnamefont{A.}~\bibnamefont{Dhar}}, \bibnamefont{and}
  \bibinfo{author}{\bibfnamefont{J.~L.} \bibnamefont{Lebowitz}},
  \bibinfo{journal}{Phys. Rev. E} \textbf{\bibinfo{volume}{85}},
  \bibinfo{pages}{041118} (\bibinfo{year}{2012}),
  \eprint{doi:10.1103/PhysRevE.85.041118}.

\bibitem[{\citenamefont{Sabhapandit}(2012)}]{Sabhapandit2012}
\bibinfo{author}{\bibfnamefont{S.}~\bibnamefont{Sabhapandit}},
  \bibinfo{journal}{Phys. Rev. E} \textbf{\bibinfo{volume}{85}},
  \bibinfo{pages}{021108} (\bibinfo{year}{2012}),
  \eprint{doi:10.1103/PhysRevE.85.021108}.

\bibitem[{\citenamefont{Dhar and Dandekar}(2015)}]{Dhar2015}
\bibinfo{author}{\bibfnamefont{A.}~\bibnamefont{Dhar}} \bibnamefont{and}
  \bibinfo{author}{\bibfnamefont{R.}~\bibnamefont{Dandekar}},
  \bibinfo{journal}{Physica A} \textbf{\bibinfo{volume}{418}},
  \bibinfo{pages}{49 } (\bibinfo{year}{2015}),
  \eprint{doi:10.1016/j.physa.2014.06.002}.

\bibitem[{\citenamefont{Velizhanin et~al.}(2015)\citenamefont{Velizhanin, Sahu,
  Chien, Dubi, and Zwolak}}]{Velizhanin2015}
\bibinfo{author}{\bibfnamefont{K.~A.} \bibnamefont{Velizhanin}},
  \bibinfo{author}{\bibfnamefont{S.}~\bibnamefont{Sahu}},
  \bibinfo{author}{\bibfnamefont{C.-C.} \bibnamefont{Chien}},
  \bibinfo{author}{\bibfnamefont{Y.}~\bibnamefont{Dubi}}, \bibnamefont{and}
  \bibinfo{author}{\bibfnamefont{M.}~\bibnamefont{Zwolak}},
  \bibinfo{journal}{Sci. Rep.} \textbf{\bibinfo{volume}{5}}
  (\bibinfo{year}{2015}), \eprint{doi:10.1038/srep17506}.

\bibitem[{\citenamefont{Murashita and Esposito}(2016)}]{Esposito2016}
\bibinfo{author}{\bibfnamefont{Y.}~\bibnamefont{Murashita}} \bibnamefont{and}
  \bibinfo{author}{\bibfnamefont{M.}~\bibnamefont{Esposito}},
  \bibinfo{journal}{Phys. Rev. E} \textbf{\bibinfo{volume}{94}},
  \bibinfo{pages}{062148} (\bibinfo{year}{2016}),
  \eprint{doi:10.1103/PhysRevE.94.062148}.

\bibitem[{\citenamefont{Craven and Nitzan}(2016)}]{craven16c}
\bibinfo{author}{\bibfnamefont{G.~T.} \bibnamefont{Craven}} \bibnamefont{and}
  \bibinfo{author}{\bibfnamefont{A.}~\bibnamefont{Nitzan}},
  \bibinfo{journal}{Proc. Natl. Acad. Sci.} \textbf{\bibinfo{volume}{113}},
  \bibinfo{pages}{9421} (\bibinfo{year}{2016}),
  \eprint{doi:10.1073/pnas.1609141113}.

\bibitem[{\citenamefont{Craven and Nitzan}(2017{\natexlab{a}})}]{craven17a}
\bibinfo{author}{\bibfnamefont{G.~T.} \bibnamefont{Craven}} \bibnamefont{and}
  \bibinfo{author}{\bibfnamefont{A.}~\bibnamefont{Nitzan}},
  \bibinfo{journal}{J. Chem. Phys.} \textbf{\bibinfo{volume}{146}},
  \bibinfo{pages}{092305} (\bibinfo{year}{2017}{\natexlab{a}}),
  \eprint{doi:10.1063/1.4971293}.

\bibitem[{\citenamefont{Craven and Nitzan}(2017{\natexlab{b}})}]{craven17b}
\bibinfo{author}{\bibfnamefont{G.~T.} \bibnamefont{Craven}} \bibnamefont{and}
  \bibinfo{author}{\bibfnamefont{A.}~\bibnamefont{Nitzan}},
  \bibinfo{journal}{Phys. Rev. Lett.} \textbf{\bibinfo{volume}{118}},
  \bibinfo{pages}{207201} (\bibinfo{year}{2017}{\natexlab{b}}),
  \eprint{doi:10.1103/PhysRevLett.118.207201}.

\bibitem[{\citenamefont{Chen et~al.}(2017)\citenamefont{Chen, Craven, and
  Nitzan}}]{craven17e}
\bibinfo{author}{\bibfnamefont{R.}~\bibnamefont{Chen}},
  \bibinfo{author}{\bibfnamefont{G.~T.} \bibnamefont{Craven}},
  \bibnamefont{and} \bibinfo{author}{\bibfnamefont{A.}~\bibnamefont{Nitzan}},
  \bibinfo{journal}{J. Chem. Phys.} \textbf{\bibinfo{volume}{147}},
  \bibinfo{pages}{124101} (\bibinfo{year}{2017}),
  \eprint{doi:10.1063/1.4990410}.

\bibitem[{\citenamefont{Marcus}(1956)}]{Marcus1956}
\bibinfo{author}{\bibfnamefont{R.~A.} \bibnamefont{Marcus}},
  \bibinfo{journal}{J. Chem. Phys.} \textbf{\bibinfo{volume}{24}},
  \bibinfo{pages}{966} (\bibinfo{year}{1956}), \eprint{doi:10.1063/1.1742723}.

\bibitem[{\citenamefont{Marcus}(1964)}]{Marcus1964}
\bibinfo{author}{\bibfnamefont{R.~A.} \bibnamefont{Marcus}},
  \bibinfo{journal}{Annu. Rev. Phys. Chem.} \textbf{\bibinfo{volume}{15}},
  \bibinfo{pages}{155} (\bibinfo{year}{1964}),
  \eprint{doi:10.1146/annurev.pc.15.100164.001103}.

\bibitem[{\citenamefont{Marcus and Sutin}(1985)}]{Marcus1985}
\bibinfo{author}{\bibfnamefont{R.~A.} \bibnamefont{Marcus}} \bibnamefont{and}
  \bibinfo{author}{\bibfnamefont{N.}~\bibnamefont{Sutin}},
  \bibinfo{journal}{Biochim. Biophys. Acta} \textbf{\bibinfo{volume}{811}},
  \bibinfo{pages}{265 } (\bibinfo{year}{1985}),
  \eprint{doi:10.1016/0304-4173(85)90014-X}.

\bibitem[{\citenamefont{Marcus}(1993)}]{Marcus1993}
\bibinfo{author}{\bibfnamefont{R.~A.} \bibnamefont{Marcus}},
  \bibinfo{journal}{Rev. Mod. Phys.} \textbf{\bibinfo{volume}{65}},
  \bibinfo{pages}{599} (\bibinfo{year}{1993}),
  \eprint{doi:10.1103/RevModPhys.65.599}.

\bibitem[{\citenamefont{H{\"a}nggi et~al.}(1990)\citenamefont{H{\"a}nggi,
  Talkner, and Borkovec}}]{rmp90}
\bibinfo{author}{\bibfnamefont{P.}~\bibnamefont{H{\"a}nggi}},
  \bibinfo{author}{\bibfnamefont{P.}~\bibnamefont{Talkner}}, \bibnamefont{and}
  \bibinfo{author}{\bibfnamefont{M.}~\bibnamefont{Borkovec}},
  \bibinfo{journal}{Rev. Mod. Phys.} \textbf{\bibinfo{volume}{62}},
  \bibinfo{pages}{251} (\bibinfo{year}{1990}),
  \eprint{10.1103/RevModPhys.62.251}.

\bibitem[{\citenamefont{Truhlar et~al.}(1996)\citenamefont{Truhlar, Garrett,
  and Klippenstein}}]{truh96}
\bibinfo{author}{\bibfnamefont{D.~G.} \bibnamefont{Truhlar}},
  \bibinfo{author}{\bibfnamefont{B.~C.} \bibnamefont{Garrett}},
  \bibnamefont{and} \bibinfo{author}{\bibfnamefont{S.~J.}
  \bibnamefont{Klippenstein}}, \bibinfo{journal}{J. Phys. Chem.}
  \textbf{\bibinfo{volume}{100}}, \bibinfo{pages}{12771}
  (\bibinfo{year}{1996}).

\bibitem[{\citenamefont{Komatsuzaki and Berry}(2001)}]{Komatsuzaki2001}
\bibinfo{author}{\bibfnamefont{T.}~\bibnamefont{Komatsuzaki}} \bibnamefont{and}
  \bibinfo{author}{\bibfnamefont{R.~S.} \bibnamefont{Berry}},
  \bibinfo{journal}{Proc. Natl. Acad. Sci.} \textbf{\bibinfo{volume}{98}},
  \bibinfo{pages}{7666} (\bibinfo{year}{2001}),
  \eprint{doi:10.1073/pnas.131627698}.

\bibitem[{\citenamefont{Bartsch et~al.}(2005)\citenamefont{Bartsch, Hernandez,
  and Uzer}}]{dawn05a}
\bibinfo{author}{\bibfnamefont{T.}~\bibnamefont{Bartsch}},
  \bibinfo{author}{\bibfnamefont{R.}~\bibnamefont{Hernandez}},
  \bibnamefont{and} \bibinfo{author}{\bibfnamefont{T.}~\bibnamefont{Uzer}},
  \bibinfo{journal}{Phys. Rev. Lett.} \textbf{\bibinfo{volume}{95}},
  \bibinfo{pages}{058301(1)} (\bibinfo{year}{2005}),
  \eprint{doi:10.1103/PhysRevLett.95.058301}.

\bibitem[{\citenamefont{Nitzan}(2006)}]{Nitzan2006chemical}
\bibinfo{author}{\bibfnamefont{A.}~\bibnamefont{Nitzan}},
  \emph{\bibinfo{title}{Chemical Dynamics in Condensed Phases: Relaxation,
  Transfer, and Reactions in Condensed Molecular Systems}}
  (\bibinfo{publisher}{Oxford University Press}, \bibinfo{year}{2006}).

\bibitem[{\citenamefont{Hernandez et~al.}(2010)\citenamefont{Hernandez,
  Bartsch, and Uzer}}]{hern10a}
\bibinfo{author}{\bibfnamefont{R.}~\bibnamefont{Hernandez}},
  \bibinfo{author}{\bibfnamefont{T.}~\bibnamefont{Bartsch}}, \bibnamefont{and}
  \bibinfo{author}{\bibfnamefont{T.}~\bibnamefont{Uzer}},
  \bibinfo{journal}{Chem. Phys.} \textbf{\bibinfo{volume}{370}},
  \bibinfo{pages}{270} (\bibinfo{year}{2010}),
  \eprint{doi:10.1016/j.chemphys.2010.01.016}.

\bibitem[{\citenamefont{Peters}(2015)}]{Peters2015}
\bibinfo{author}{\bibfnamefont{B.}~\bibnamefont{Peters}}, \bibinfo{journal}{J.
  Phys. Chem. B} \textbf{\bibinfo{volume}{119}}, \bibinfo{pages}{6349}
  (\bibinfo{year}{2015}), \eprint{doi:10.1021/acs.jpcb.5b02547}.

\bibitem[{\citenamefont{Craven and Hernandez}(2015)}]{craven15c}
\bibinfo{author}{\bibfnamefont{G.~T.} \bibnamefont{Craven}} \bibnamefont{and}
  \bibinfo{author}{\bibfnamefont{R.}~\bibnamefont{Hernandez}},
  \bibinfo{journal}{Phys. Rev. Lett.} \textbf{\bibinfo{volume}{115}},
  \bibinfo{pages}{148301} (\bibinfo{year}{2015}),
  \eprint{doi:10.1103/PhysRevLett.115.148301}.

\bibitem[{\citenamefont{Chandler}(1987)}]{Chandler1987statmech}
\bibinfo{author}{\bibfnamefont{D.}~\bibnamefont{Chandler}},
  \emph{\bibinfo{title}{Introduction to Modern Statistical Mechanics}}
  (\bibinfo{publisher}{Oxford University Press}, \bibinfo{year}{1987}).

\bibitem[{\citenamefont{Greene}(2002)}]{Greene2002econo}
\bibinfo{author}{\bibfnamefont{W.~H.} \bibnamefont{Greene}},
  \emph{\bibinfo{title}{Econometric Analysis}} (\bibinfo{publisher}{Prentice
  Hall}, \bibinfo{year}{2002}).

\bibitem[{\citenamefont{Ranganatham}(2006)}]{Ranganatham2006}
\bibinfo{author}{\bibfnamefont{M.}~\bibnamefont{Ranganatham}},
  \emph{\bibinfo{title}{Investment Analysis and Portfolio Management}}
  (\bibinfo{publisher}{Pearson Education India}, \bibinfo{year}{2006}).

\bibitem[{\citenamefont{Reilly and Brown}(2011)}]{Reilly2011}
\bibinfo{author}{\bibfnamefont{F.~K.} \bibnamefont{Reilly}} \bibnamefont{and}
  \bibinfo{author}{\bibfnamefont{K.~C.} \bibnamefont{Brown}},
  \emph{\bibinfo{title}{Investment Analysis and Portfolio Management}}
  (\bibinfo{publisher}{Cengage Learning}, \bibinfo{year}{2011}).

\bibitem[{\citenamefont{Sortino and Van Der~Meer}(1991)}]{Sortino1991}
\bibinfo{author}{\bibfnamefont{F.~A.} \bibnamefont{Sortino}} \bibnamefont{and}
  \bibinfo{author}{\bibfnamefont{R.}~\bibnamefont{Van Der~Meer}},
  \bibinfo{journal}{J. Portfolio Manage.} \textbf{\bibinfo{volume}{17}},
  \bibinfo{pages}{27} (\bibinfo{year}{1991}).

\bibitem[{\citenamefont{Sortino and Price}(1994)}]{Sortino1994}
\bibinfo{author}{\bibfnamefont{F.~A.} \bibnamefont{Sortino}} \bibnamefont{and}
  \bibinfo{author}{\bibfnamefont{L.~N.} \bibnamefont{Price}},
  \bibinfo{journal}{J. Invest.} \textbf{\bibinfo{volume}{3}},
  \bibinfo{pages}{59} (\bibinfo{year}{1994}).

\bibitem[{\citenamefont{Keating and Shadwick}(2002)}]{Keating2002universal}
\bibinfo{author}{\bibfnamefont{C.}~\bibnamefont{Keating}} \bibnamefont{and}
  \bibinfo{author}{\bibfnamefont{W.~F.} \bibnamefont{Shadwick}},
  \bibinfo{journal}{J. Perf. Measure.} \textbf{\bibinfo{volume}{6}},
  \bibinfo{pages}{59} (\bibinfo{year}{2002}).

\bibitem[{\citenamefont{Ang et~al.}(2006)\citenamefont{Ang, Chen, and
  Xing}}]{Ang2006downside}
\bibinfo{author}{\bibfnamefont{A.}~\bibnamefont{Ang}},
  \bibinfo{author}{\bibfnamefont{J.}~\bibnamefont{Chen}}, \bibnamefont{and}
  \bibinfo{author}{\bibfnamefont{Y.}~\bibnamefont{Xing}},
  \bibinfo{journal}{Rev. Financ. Stud.} \textbf{\bibinfo{volume}{19}},
  \bibinfo{pages}{1191} (\bibinfo{year}{2006}),
  \eprint{doi:10.1093/rfs/hhj035}.

\bibitem[{\citenamefont{Zwanzig}(2001)}]{zwan01book}
\bibinfo{author}{\bibfnamefont{R.}~\bibnamefont{Zwanzig}},
  \emph{\bibinfo{title}{Nonequilibrium Statistical Mechanics}}
  (\bibinfo{publisher}{Oxford University Press}, \bibinfo{address}{London},
  \bibinfo{year}{2001}).

\bibitem[{\citenamefont{Revuelta et~al.}(2017)\citenamefont{Revuelta, Craven,
  Bartsch, and Hernandez}}]{craven17c}
\bibinfo{author}{\bibfnamefont{F.}~\bibnamefont{Revuelta}},
  \bibinfo{author}{\bibfnamefont{G.~T.} \bibnamefont{Craven}},
  \bibinfo{author}{\bibfnamefont{T.}~\bibnamefont{Bartsch}}, \bibnamefont{and}
  \bibinfo{author}{\bibfnamefont{R.}~\bibnamefont{Hernandez}},
  \bibinfo{journal}{J. Chem. Phys.} \textbf{\bibinfo{volume}{147}},
  \bibinfo{pages}{074104} (\bibinfo{year}{2017}),
  \eprint{doi:10.1063/1.4997571}.

\bibitem[{\citenamefont{Craven et~al.}(2017)\citenamefont{Craven, Junginger,
  and Hernandez}}]{craven17d}
\bibinfo{author}{\bibfnamefont{G.~T.} \bibnamefont{Craven}},
  \bibinfo{author}{\bibfnamefont{A.}~\bibnamefont{Junginger}},
  \bibnamefont{and}
  \bibinfo{author}{\bibfnamefont{R.}~\bibnamefont{Hernandez}},
  \bibinfo{journal}{Phys. Rev. E} \textbf{\bibinfo{volume}{96}},
  \bibinfo{pages}{022222} (\bibinfo{year}{2017}),
  \eprint{doi:10.1103/PhysRevE.96.022222}.

\bibitem[{\citenamefont{Chandrasekhar}(1943)}]{Chandrasekhar1943}
\bibinfo{author}{\bibfnamefont{S.}~\bibnamefont{Chandrasekhar}},
  \bibinfo{journal}{Rev. Mod. Phys.} \textbf{\bibinfo{volume}{15}},
  \bibinfo{pages}{1} (\bibinfo{year}{1943}),
  \eprint{doi:10.1103/RevModPhys.15.1}.

\bibitem[{\citenamefont{Uhlenbeck and Ornstein}(1930)}]{OrnsteinUhlenbeck1930}
\bibinfo{author}{\bibfnamefont{G.~E.} \bibnamefont{Uhlenbeck}}
  \bibnamefont{and} \bibinfo{author}{\bibfnamefont{L.~S.}
  \bibnamefont{Ornstein}}, \bibinfo{journal}{Phys. Rev.}
  \textbf{\bibinfo{volume}{36}}, \bibinfo{pages}{823} (\bibinfo{year}{1930}),
  \eprint{doi:10.1103/PhysRev.36.823}.

\bibitem[{\citenamefont{Matyushov}(2016)}]{matyushov16c}
\bibinfo{author}{\bibfnamefont{D.~V.} \bibnamefont{Matyushov}},
  \bibinfo{journal}{Proc. Natl. Acad. Sci.} \textbf{\bibinfo{volume}{113}},
  \bibinfo{pages}{9401} (\bibinfo{year}{2016}),
  \eprint{doi:10.1073/pnas.1610542113}.

\bibitem[{\citenamefont{Cohen}(2015)}]{Cohen2015review}
\bibinfo{author}{\bibfnamefont{L.}~\bibnamefont{Cohen}}, in
  \emph{\bibinfo{booktitle}{Mathematical Analysis, Probability and Applications
  - Plenary Lectures ISAAC 2015}} (\bibinfo{organization}{Springer},
  \bibinfo{year}{2015}), pp. \bibinfo{pages}{1--35}.

\bibitem[{not({\natexlab{a}})}]{note1}
\bibinfo{note}{Results for the different case $E^\ddag = E(0)$ and $\rho_0 =
  \delta(v-v_0)$ are given in the Supplementary Material.}

\bibitem[{not({\natexlab{b}})}]{note2}
\bibinfo{note}{Derivations of restricted transport properties are given in the
  Supplementary Material.}

\bibitem[{\citenamefont{Ng and Geller}(1969)}]{Ng1969}
\bibinfo{author}{\bibfnamefont{E.~W.} \bibnamefont{Ng}} \bibnamefont{and}
  \bibinfo{author}{\bibfnamefont{M.}~\bibnamefont{Geller}},
  \bibinfo{journal}{Journal of Research of the National Bureau of Standards B}
  \textbf{\bibinfo{volume}{73}}, \bibinfo{pages}{1–20}
  (\bibinfo{year}{1969}).

\bibitem[{\citenamefont{Simon and Divsalar}(1998)}]{Divsalar1998}
\bibinfo{author}{\bibfnamefont{M.~K.} \bibnamefont{Simon}} \bibnamefont{and}
  \bibinfo{author}{\bibfnamefont{D.}~\bibnamefont{Divsalar}},
  \bibinfo{journal}{IEEE Trans. Commun.} \textbf{\bibinfo{volume}{46}},
  \bibinfo{pages}{200} (\bibinfo{year}{1998}), \eprint{doi:10.1109/26.659479}.

\bibitem[{\citenamefont{Fayed and Atiya}(2014)}]{Fayed2014}
\bibinfo{author}{\bibfnamefont{H.}~\bibnamefont{Fayed}} \bibnamefont{and}
  \bibinfo{author}{\bibfnamefont{A.}~\bibnamefont{Atiya}},
  \bibinfo{journal}{Math. Comp} \textbf{\bibinfo{volume}{83}},
  \bibinfo{pages}{235} (\bibinfo{year}{2014}),
  \eprint{doi:10.1090/S0025-5718-2013-02720-2}.

\bibitem[{\citenamefont{Fayed et~al.}(2015)\citenamefont{Fayed, Atiya, and
  Badawi}}]{Fayed2015}
\bibinfo{author}{\bibfnamefont{H.}~\bibnamefont{Fayed}},
  \bibinfo{author}{\bibfnamefont{A.}~\bibnamefont{Atiya}}, \bibnamefont{and}
  \bibinfo{author}{\bibfnamefont{A.}~\bibnamefont{Badawi}},
  \bibinfo{journal}{Math. Sci. Lett.} \textbf{\bibinfo{volume}{4}},
  \bibinfo{pages}{249} (\bibinfo{year}{2015}),
  \eprint{doi:10.12785/msl/040305}.

\end{thebibliography}
\end{document}